\documentclass[prapplied,superscriptaddress,notitlepage,twocolumn]{revtex4-2}

\usepackage{soul}
\usepackage{amsthm}
\usepackage{amsmath}
\usepackage{amssymb}
\usepackage{graphicx}
\usepackage{epstopdf}
\usepackage{subfig}
\usepackage{multirow}
\usepackage{url}
\usepackage{float}
\usepackage{bm}
\usepackage{ifthen}
\usepackage[usenames,dvipsnames]{color}
\usepackage{mathrsfs}
\usepackage[colorlinks=false,citecolor=blue,urlcolor=black]{hyperref}

\newcommand{\Tr}{\mathrm{ Tr }}

\newcommand{\be}{\begin{equation}}
\newcommand{\ee}{\end{equation}}

\usepackage[absolute]{textpos}
\usepackage{xcolor}

\usepackage{placeins}

\newcommand{\sket}[1]{{\ensuremath{\lvert#1\rangle}}}
\newcommand{\lket}[1]{{\ensuremath{\left\lvert#1\right\rangle}}}
\newcommand{\ket}[1]{\if@display\lket{#1}\else\sket{#1}\fi}

\newcommand{\sbra}[1]{{\ensuremath{\langle#1\rvert}}}
\newcommand{\lbra}[1]{{\ensuremath{\left\langle#1\right\rvert}}}
\newcommand{\bra}[1]{\if@display\lbra{#1}\else\sbra{#1}\fi}

\newcommand{\sbraket}[2]{{\ensuremath{\langle#1\rvert#2\rangle}}}
\newcommand{\lbraket}[2]{{\ensuremath{\left\langle#1\!\left\rvert\vphantom{#1}#2\right.\!\right\rangle}}}
\newcommand{\braket}[2]{\if@display\lbraket{#1}{#2}\else\sbraket{#1}{#2}\fi}

\newcommand{\sketbra}[2]{{\ensuremath{\lvert #1\rangle\!\langle #2\rvert}}}
\newcommand{\lketbra}[2]{{\ensuremath{\left\lvert #1\right\rangle\!\!\left\langle #2\right\rvert}}}
\newcommand{\ketbra}[2]{\if@display\lketbra{#1}{#2}\else\sketbra{#1}{#2}\fi}

\def\by{\mathbf{y}}
\def\bx{\mathbf{x}}

\theoremstyle{plain}

\theoremstyle{definition}

\begin{document}
\title{Coexistent quantum channel characterization using spectrally resolved Bayesian quantum process tomography}
\author{Joseph C. Chapman}
\email{chapmanjc@ornl.gov}
\affiliation{Quantum Information Science Section, Oak Ridge National Laboratory, Oak Ridge, TN 37831}
\author{Joseph M. Lukens}
\affiliation{Quantum Information Science Section, Oak Ridge National Laboratory, Oak Ridge, TN 37831}
\affiliation{Research Technology Office, Arizona State University, Tempe, AZ 85287}
\author{Muneer Alshowkan}
\affiliation{Quantum Information Science Section, Oak Ridge National Laboratory, Oak Ridge, TN 37831}
\author{Nageswara Rao}
\affiliation{Advanced Computing Methods for Engineered Systems Section, Oak Ridge National Laboratory, Oak Ridge, TN 37831}
\author{Brian T. Kirby}
\affiliation{Tulane University, New Orleans, LA 70118}
\affiliation{DEVCOM Army Research Laboratory, Adelphi, MD 20783}
\author{Nicholas A. Peters}
\affiliation{Quantum Information Science Section, Oak Ridge National Laboratory, Oak Ridge, TN 37831}
\begin{abstract}
The coexistence of quantum and classical signals over the same optical fiber with minimal degradation of the transmitted quantum information is critical for operating large-scale quantum networks over the existing communications infrastructure. 
Here, we systematically characterize the quantum channel that results from simultaneously distributing approximate single-photon polarization-encoded qubits and classical light of varying intensities through fiber-optic channels of up to 15~km.
Using spectrally resolved quantum process tomography with a Bayesian reconstruction method we developed, we estimate the full quantum channel from experimental photon counting data, both with and without classical background. Furthermore, although we find the exact channel description to be a weak function of the pump polarization, we nevertheless show that the coexistent fiber-based quantum channel has high process fidelity with an ideal depolarizing channel when the noise is dominated by Raman scattering. These results provide a basis for the future development of quantum repeater designs and quantum error correcting codes for real-world channels and inform models used in the analysis and simulation of quantum networks. 
\end{abstract}
\maketitle


\section{Motivation and Background}
The telecommunications industry has deployed a worldwide infrastructure of fiber-optic links and hierarchical networks. Nevertheless, most quantum communications experiments have been performed in laboratories or using so-called ``dark'' fiber that is devoid of light from classical network traffic and the noise it generates. For quantum communications to leverage a deployed fiber-optic network without disruption, coexistence with the classical networking traffic that dominates the fiber is essential~\cite{el_19970147}.
To date, experimental characterization of the noise produced by conventional fiber-optic communication has largely been focused on how the noise from classical signals affects quantum key distribution~\cite{Peters_2009,Chapuran_2009,Eraerds_2010,PhysRevX.2.041010,6819012,Frohlich2015,Aleksic:15,PhysRevA.95.012301,Valivarthi_2019,doi:10.1063/5.0060232,Tessinari:21,Berrevoets2022}. While some have included spectral measurements of the noise at the single-photon level~\cite{Peters_2009,Chapuran_2009,Eraerds_2010,PhysRevX.2.041010,Aleksic:15,PhysRevA.95.012301,Lin:19,Valivarthi_2019,Tessinari:21}, the noise produced by conventional fiber-optic transceivers includes a number of distinct processes governed by rich and diverse physics, requiring measurements beyond spectral power density alone for full characterization. Furthermore, while some experiments have explored coexistence for entanglement distribution using visibility measurements~\cite{1637099,Sauge:07,Yuan:2019, Thomas2022}, they did not analyze the noise based on quantum process tomography (QPT)---the standard approach for obtaining a complete characterization of an arbitrary quantum channel~\cite{Poyatos1997,Chuang1997}.

In this work, we utilize Bayesian inference to fully characterize the quantum process experienced by a single polarization qubit in optical fiber, for a variety of classical power levels and up to channel lengths of 15.5~km. We also employ spectrally resolved polarization quantum state tomography (QST) to characterize the noise alone. Finally, we apply these results to inform a working model for quantum network design that 
can be used to establish spectro-temporal filtering guidelines supporting a desired quantum channel fidelity.
Importantly, our measurements show that the quantum channel in a fiber-optic link---when dominated by Raman scattering noise---is depolarizing to a close approximation, with minor dependence on the degree of polarization of the Raman pump. 
These results inform error models of realistic  quantum channels that can be employed for optimizing future quantum repeaters and quantum error correcting codes, as well as simulations of quantum networks and estimations of their capacities.

We begin by introducing our methods for Bayesian QPT in Sec.~\ref{section:BQPT}, and describing our experimental setup in Sec.~\ref{section:expsetup}. Then in Sec.~\ref{section:results}, we show the results of our measurements and compare them with known channel models before concluding our findings in Sec.~\ref{section:conclusion}. For the interested reader, in Appendix~\ref{App:QPT} we describe our Bayesian quantum process tomography calculations in further detail. In Appendix~\ref{App:Chsim}, we simulate several channel models and compare the results of our Bayesian QPT workflow to the ground truth. Finally, in Appendix~\ref{App:QPTdata}, we provide the rest of the measurement results not included in Sec.~\ref{section:results}.

\section{Bayesian Quantum Process Tomography} \label{section:BQPT}
In this section, we discuss our method for QPT analysis using Bayesian inference. 
A generic quantum channel can be represented in terms of Kraus operators $A_k$ as
\begin{equation}
\label{eq:Kraus}
    \mathcal{E}(\rho)=\sum_{k=1}^{K}{A_k\rho A_k^{\dagger}},
\end{equation}
where, for $D$-dimensional input/output Hilbert spaces, $K\leq D^2$ defines the Choi rank. Following Ref.~\cite{doi:10.1063/5.0038838}, we use the complex numbers $y_j$ for $j \in \{1,...,K D^2\}$ to parameterize the Kraus operators. Each of the $K$ nonoverlapping length-$D^2$ subsets can be used to populate a $D\times D$ matrix $G_k$, from which we define
\begin{equation}
    A_k=G_k H^{-1/2},
\end{equation}
where $H\equiv\sum_{k=1}^K {G_k^{\dagger}G_k}$ to ensure that $\sum_{k=1}^K A_k^\dagger A_k = I_D$ (the $D\times D$ identity)---i.e., the operation is trace-preserving. In a practical experiment with non-unit photon transmission, one can postselect on the cases where a photon is detected. In this subspace, the channel is trace-preserving~\cite{PhysRevLett.90.193601,PhysRevLett.93.080502}. Furthermore, in the presence of  incoherent added noise---which we consider below---postselection also maintains a trace-preserving channel, where the incoherent added noise appears in the output state as a mixture with the input state.

As shown in Ref.~\cite{doi:10.1063/5.0038838}, for  $K=D^2$ and $y_j$ chosen as independent and identically distributed complex normal random variables [$y_j \sim \mathcal{N}(0,1)+i\mathcal{N}(0,1)$], the quantum channel distribution is the Lebesgue (flat) measure and is thus a suitable uniform prior. For Markov chain Monte Carlo (MCMC) sampling, this formulation is particularly convenient because it contains only normally distributed parameters in the prior (a total of $2D^4$ real numbers).

For QPT, we consider the case where $L$ input states $\ket{\phi_l}$ are sent into the system, and projective measurements on $J$ output states $\ket{\psi_j}$ are performed. Our approach can readily accommodate mixed input states $\rho_l$ as well, but we specialize to the case of pure inputs here as it reduces computational complexity while still accurately reflecting the quantum states we prepare experimentally (whose measured purities exceed 0.99). For a given input state $\ket{\phi_l}$, quantum channel (defined by Kraus operators $A_k$), and output state $\ket{\psi_j}$, we define the probability
\begin{equation}
\label{eq:plj}
    p_{l j}=\bra{\psi_j}\mathcal{E}(\ket{\phi_l}\bra{\phi_l})\ket{\psi_j}=\sum_{k=1}^K|\bra{\psi_j}A_k\ket{\phi_l}|^2.
\end{equation}
The mean number of photons detected in $\ket{\psi_j}$ in a given integration window $\tau$ with $\ket{\phi_l}$ input is therefore $\overline{N}_{lj}=\Phi \tau p_{l j} $, where $\Phi$ is the incident flux (in units of photons per unit time). We have chosen to prepare and measure an overcomplete set of states, namely $\ket{0}\equiv\ket{H}$, $\ket{1}\equiv\ket{V}$, $\ket{\pm}=(\ket{0}\pm\ket{1})/\sqrt{2}$, and  $\ket{\pm i} = (\ket{0}-i\ket{1})/\sqrt{2}$, where $\ket{H}$ ($\ket{V}$) represents a horizontally (vertically) polarized single-photon state.

To reconstruct the quantum channel, we use Bayes' theorem and a particularly efficient MCMC method, known as ``preconditioned Crank--Nicolson'' (pCN)~\cite{Cotter2013}, which we previously introduced in the context of QST~\cite{Lukens2020b} and have since applied to a variety of inference examples~\cite{Simmerman2020, Lu2020b, Lohani2021, Lu2022, Chapman2022, Lu2022b}; pCN returns samples of the length-$D^4$ vector $\by=(y_1,...,y_{D^4})$ that parametrizes the quantum channel. For more information on implementing this algorithm, see Appendix~\ref{App:QPT}. For examples of our method applied to noiseless simulated data for the depolarizing, dephasing, and bit-flip channels, see Appendix~\ref{App:Chsim}, where we find excellent agreement between the ground truth and Bayesian inference results. We note that Bayesian QPT has been explored previously, both in simulation~\cite{Granade2016} and experiment~\cite{Pogorelov2017}. However, our approach builds on these formulations via (i)~application of pCN sampling techniques and (ii)~adoption of the Kraus operator parametrization introduced in Ref.~\cite{doi:10.1063/5.0038838}, where it was recommended as being more computationally efficient than the Choi~\cite{Granade2016} and Stinespring~\cite{Pogorelov2017} constructions enlisted in earlier Bayesian contexts.

\section{Experimental Setup}\label{section:expsetup}
\begin{figure*}[tb!]
\begin{center}
\includegraphics[scale=0.7]{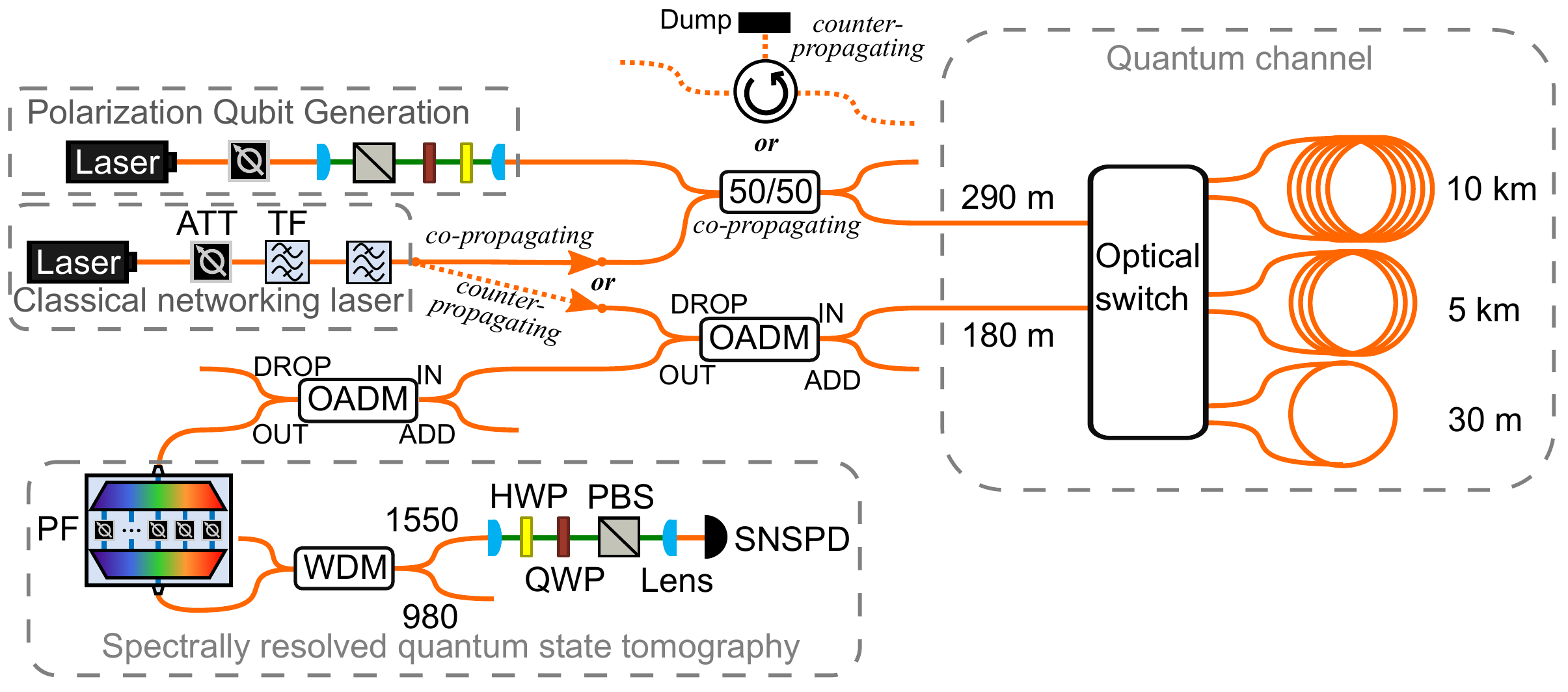}
\end{center}
\caption{Experimental setup for quantum and classical signals co-propagating and counter-propagating (dotted lines). The quantum state is generated using an attenuated laser and free-space polarization optics before being sent through the quantum channel. The classical networking laser is prepared and coupled into the channel carrying the the quantum signal. In the counter-propagating configuration, 
the output classical light is removed with a fiber-optic circulator in place of the 50/50 fused coupler. We use notch filtering to remove the majority of the classical networking laser power at the receiver. The rest of the spectrum is sent to the spectrally resolved QST subsystem for measurement. Orange (green) lines represent telecom single-mode fiber (free-space propagation). ATT $\equiv$ optical attenuator(s). HWP $\equiv$ half-wave plate. OADM $\equiv$ optical add drop multiplexer (notch filter). PBS $\equiv$ polarizing beamsplitter. PF $\equiv$ programmable filter. QWP $\equiv$ quarter-wave plate. SNSPD $\equiv$ superconducting-nanowire single-photon detector. TF $\equiv$ tunable bandpass filter. WDM $\equiv$ wavelength-division multiplexer.}
\label{fig:setup}
\end{figure*}

Fig.~\ref{fig:setup} provides a schematic of the experimental test setup. To generate polarization qubits, we use attenuated coherent states from a tunable continuous-wave (CW) laser (Pure Photonics PPCL550) as an approximate single-photon source ($<$10~Mcps (megacounts per second) before entering the quantum channel). Then, we prepare the polarization state in free space using a polarizing beamsplitter (Thorlabs CCM1-PBS254) followed by broadband zero-order quarter- and  half-wave plates, both in motorized rotation mounts (Thorlabs K10CR1/M). After that, we couple the light back into single-mode fiber. 
To make spectrally resolved QPT measurements across most of the telecommunications C-band, we tune the qubit laser wavelength from 1528.38--1564.68~nm at a step size of 0.4~nm (50~GHz).
Our coexisting classical light is provided by another tunable CW laser (Pure Photonics PPCL550) centered at 1542.5~nm, attenuated to launch powers of either $-$8.5 or $-$5.5~dBm into the quantum channel (depending on test and chosen to avoid saturation of our single-photon detectors and timetagger with Raman noise counts).
The pedestal of the laser (measured 70~dB down from the peak at a spectral resolution of 0.02~nm) is filtered with about 150~dB of isolation using two tunable bandpass filters (DiCon Fiberoptics TF-1550-0.8-9/9LT-FC/A-1), without which the laser pedestal would completely dominate all quantum signals on the single-photon detector.

For the co-propagating measurements (Fig.~\ref{fig:setup} with solid lines), the polarization qubits and the classical networking laser are combined using a 50/50 fused coupler (Thorlabs TW1550R5A2). At the receiver, they are separated by two cascaded fiber-optic optical add-drop multiplexers (OADM; AC Photonics OADM201C442111) which act as notch filters removing the main classical networking laser but not the noise at other wavelengths generated during propagation. For the counter-propagating measurements (Fig.~\ref{fig:setup} including dotted line modifications), the classical networking laser is introduced into the channel at the DROP port of the first OADM.
To prevent feedback into the polarization qubit generation setup, the output classical light is removed with a fiber-optic circulator (General Photonics CIR-15-SM-90-FC/APS) in place of the 50/50 fused coupler.  The counter-propagating configuration allows us to probe the effects from back-scattered noise in the same direction as the quantum signal.

The quantum channels under test include about 290~m of fiber from the lab, where the quantum and classical signals are mixed, to a networking testbed that contains an all-optical switch (Huber + Suhner Polatis 6000 16$\times$16) with connected spools of single-mode fiber of lengths 30~m, 5~km, and 10~km. The quantum signals are then directed back through a different 180-m link to the original lab for measurement. We do not implement polarization stabilization to counteract polarization drift, yet we do randomize the wavelength ordering of each scan to decorrelate potential temporal drift from birefringence across the spectrum. Similarly, within each QPT for a given wavelength channel, we randomize the polarization generation and measurement order. In the absence of coexistence-induced noise, a polarization transformation which is stable during the collection of a single-wavelength dataset (about 6~min.) can be validated \emph{ex post facto} through a high unitarity [see Fig.~\ref{fig:CPNN}(a)]; on the other hand, polarization stability on the scale of about 9~hours reveals itself through steady trends in the relative process fidelity with wavelength [see Fig.~\ref{fig:CPNN}(c)]. From the findings in the following sections, our tests generally show consistent temporal stability for all measurements, though at 15.5~km some small effects from polarization drift are noticeable.

To make spectrally resolved tomography measurements, we send the output of the OADM filters to a programmable filter (Finisar Waveshaper 1000A) applying a 0.37-nm full-width at half-maximum bandpass filter which we tune from 1528.38--1564.68~nm to match the center wavelength of the input polarization qubit. The output of the filter goes through a 980/1550-nm wavelength-division multiplexer (AC Photonics DA9855202A620511) to remove leaked diagnostic light from the programmable filter and is connected to our fiber-coupled free-space polarization analysis system which consists of the same elements as the qubit generation setup, but in reverse. 
The output is then coupled back into single-mode fiber before being sent to a telecom-optimized superconducting-nanowire single-photon detector (Quantum Opus).  
The detection events are collected over an integration time of 1~s on an in-house timetagger~\cite{Alshowkan:22} comprising a system-on-chip development board (Digilent Zedboard using Xilinx Zynq-7000 ARM/FPGA SoC), capable of reliably processing about 1~Mcps. We use MATLAB to control the quantum and classical signal lasers, motorized rotation mounts,  and programmable filter, as well as to retrieve data from the timetagger, resulting in fully automated spectrally resolved quantum state and quantum process tomographies. A full-spectrum QST (QPT) scan takes approximately  1 hour (9 hours) to complete.

\section{Results}\label{section:results}
Here we describe the  QPT measurements for several different configurations of fiber-optic channel length, quantum signal intensity, classical networking laser intensity, and propagation direction. In general, random birefringence in single-mode fiber~\cite{Agrawal2021, Gordon2000} leads to unpredictable polarization rotations, which complicates the definition of an ideal benchmark against which our observed quantum processes can be compared. Importantly, however, provided that the ground truth transformation can be modeled as unitary matrix over the band of interest and remains stable over the duration of an experiment (or feedback loop in the case of active stabilization), it can be completely compensated with a polarization controller. Accordingly, the defining mark of an ideal process is its equivalence to a unitary operation, or in other words a Choi rank of $K=1$ [Eq.~\eqref{eq:Kraus}]. But since a Choi rank $K>1$ does not account for the relative weights of the contributing operators, we enlist the more informative unitarity $U(\mathcal{E})\in[0,1]$ to quantify the ideality of our measured quantum processes. The unitarity is based on the purity of the channel output state, when averaged over all possible pure input states with the identity components subtracted~\cite{Wallman_2015}. We compute unitarity using the matrix method of Refs.~\cite{Wallman_2015,PhysRevResearch.4.023041}. 

While the specific unitary transformation applied by the fiber for a given spectral channel is largely unimportant for applications, its variation with wavelength---i.e., polarization mode dispersion (PMD)---can pose substantial deleterious impacts on broadband quantum signals~\cite{antonelli2011sudden,jones2018tuning,jones2020exploring}. Although not a major focus of the present investigation, we do characterize wavelength variation by computing the fidelity of the average Choi matrix of each channel with respect to that at the lowest wavelength (1528.38~nm), which we call relative process fidelity. Finally, in order to gauge the utility of our measured quantum processes for quantum communications in general, we compute the channel capacity~\cite{PhysRevA.55.1613,PhysRevA.54.2629}.
For more information on these calculations, see Appendix~\ref{App:QPT}.

\subsection{Co-propagating configuration}
\begin{figure}
   \centerline{\includegraphics[scale=0.6]{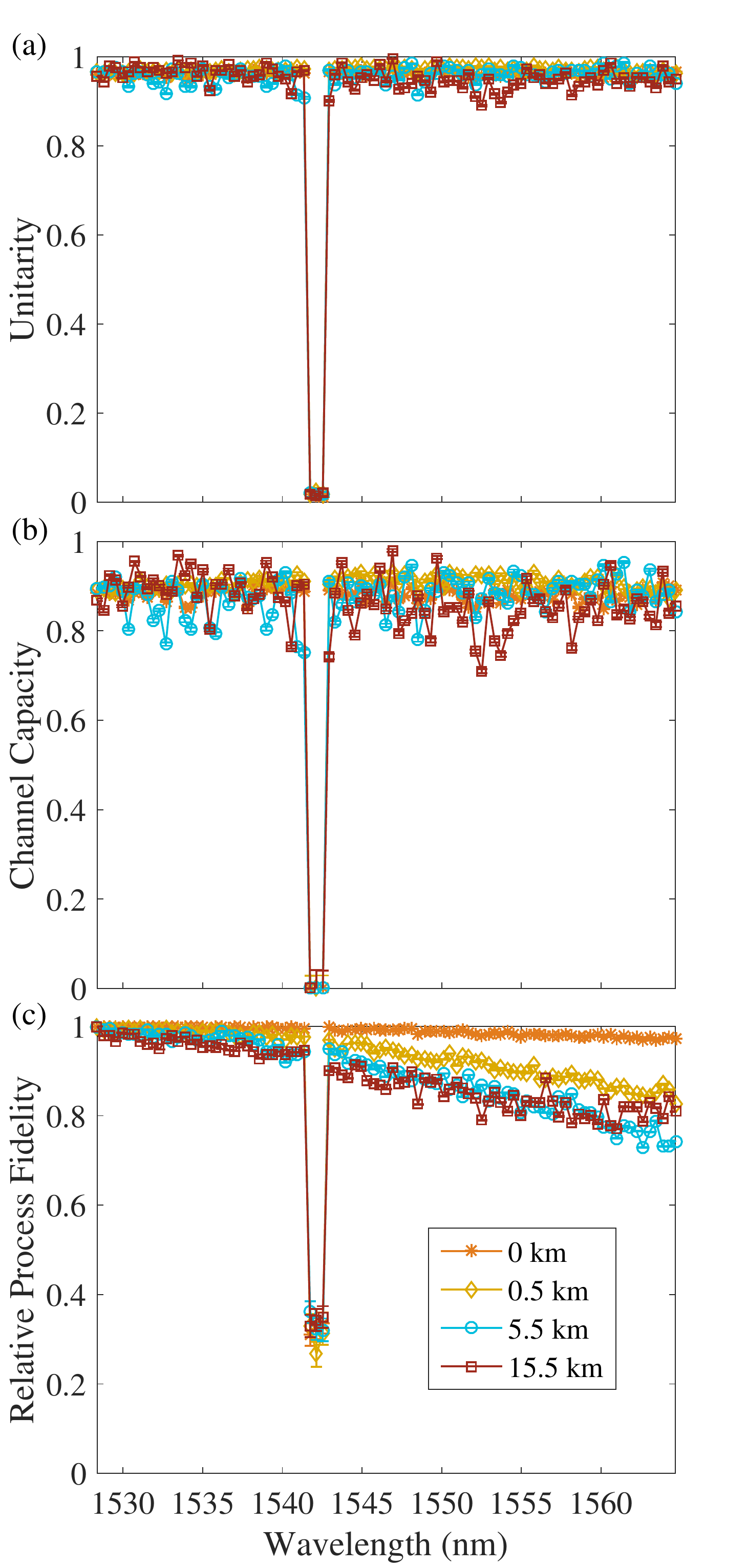}}
    \caption{Quantum process tomography versus wavelength with no added noise in co-propagating configuration. (a) Unitarity,  (b) channel capacity (qubits/channel use), and (c) relative process fidelity for several different fiber lengths. We used the Kraus operators of the quantum channel at 1528.38~nm as the reference process in (c). The dip at 1542~nm results from the notch filter.}
    \label{fig:CPNN}
\end{figure}

\begin{figure}
   \centerline{\includegraphics[scale=0.6]{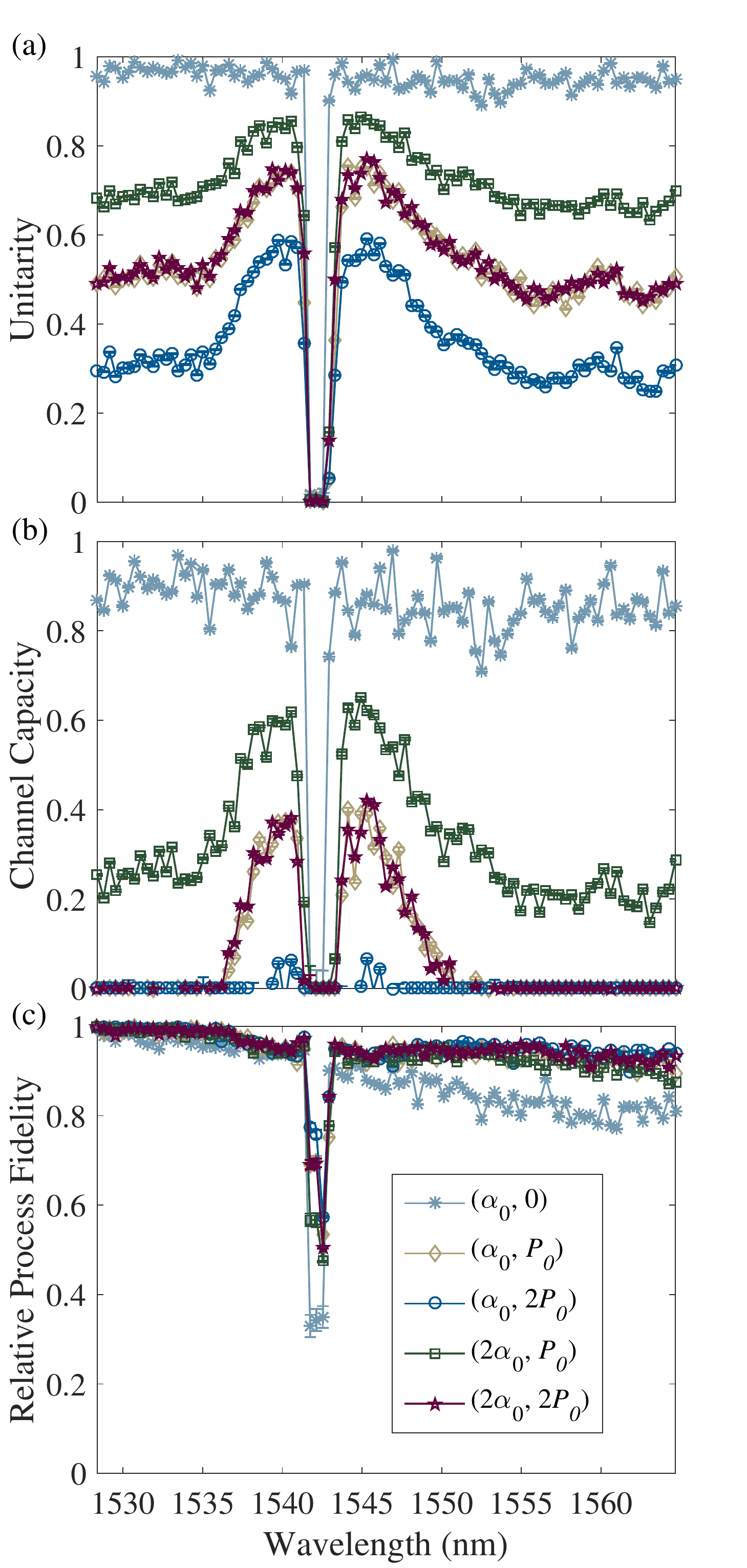}}
    \caption{Quantum process tomography at 15.5~km with added co-propagating noise.  (a) Unitarity, (b) channel capacity (qubits/channel use), and (c) relative process fidelity versus wavelength for several different combinations of quantum signal and classical noise levels.  Our base quantum signal level, $\alpha_0\approx$125 kcps detected at 0.5~km. $P_0\approx-8.5$~dBm launch power of the classical networking laser into the quantum channel. We used the Kraus operators of the quantum channel at 1528.38~nm as the reference process in (c). The dip near 1542~nm is an artifact from the notch filter.}
    \label{fig:CP15km}
\end{figure}

\begin{figure}
   \centerline{\includegraphics[scale=0.6]{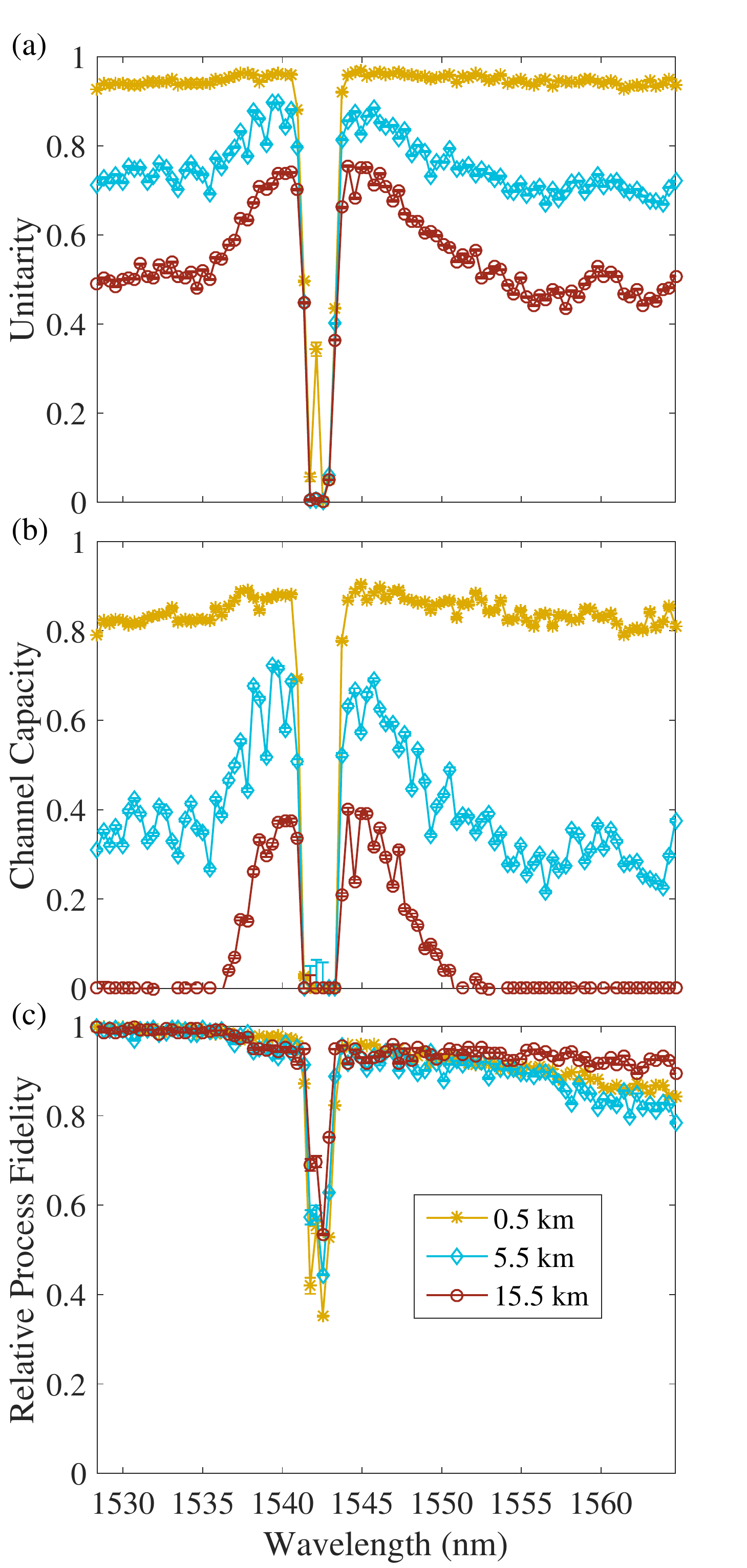}}
     \caption{Quantum process tomography versus wavelength for several distances with added co-propagating noise ($\alpha_0$,$P_0$). (a) Unitarity, (b) channel capacity (qubits/channel use), and (c) relative process fidelity for a variety of fiber lengths. Our base quantum signal level, $\alpha_0\approx$125 kcps detected at 0.5~km. $P_0\approx-8.5$~dBm launch power of the classical networking laser into the quantum channel. We used the Kraus operators of the quantum channel at 1528.38~nm as the reference process in (c). The dip near 1542~nm is an artifact from the notch filter.}
    \label{fig:CPqc912}
\end{figure}

In order to obtain baseline quantum channels, we perform  QPT scans for several different fiber lengths in the co-propagating configuration, but with the classical networking laser turned off, including a reference scan without traveling to the optical switch; the results appear in Fig.~\ref{fig:CPNN}.
The error bars are one standard deviation, calculated from 1024 samples of the MCMC chain. In all panels of Figs.~\ref{fig:CPNN}--\ref{fig:CPqc912}, the sharp drop around 1542~nm results from the notch filter used in the optical system to remove the 1542.5-nm classical networking laser. The polarization drift was slow during measurements of a single channel, so the unitarity  [Fig.~\ref{fig:CPNN}(a)] and channel capacity [Fig.~\ref{fig:CPNN}(b)] are relatively stable, apart from small fluctuations appearing in the the 15.5-km dataset. 
While one might expect that, in general, quantum networks need active polarization stabilization~\cite{Poppe2004, Xavier2008, Xavier2009, Wang2009}, there are important special cases where such tracking may not be required---e.g., short quantum local-area networks~\cite{Alshowkan2021, Alshowkan:22}, long submarine fibers~\cite{Wengerowsky2019}, or even metropolitan spans of buried fiber~\cite{Shi2020}.

Overall, we notice comparable behavior between Fig.~\ref{fig:CPNN}(a) and \ref{fig:CPNN}(b), except that the channel capacity appears to be more sensitive to imperfections than the unitarity. In Fig.~\ref{fig:CPNN}(b), the average channel capacity for the reference scan (0~km) is slightly lower than that for the 0.5-km and 5.5-km scans; the channel capacity out to 5.5~km changes very little in the case of no added noise, but there exists some polarization-dependent loss (about 0.5--1~dB) in our measurement system. We have observed that even at 0~km, the channel capacity can vary by about 0.05 qubits/channel-use depending on the specific fiber polarization transformation of the whole system at the moment. On the other hand, the relative process fidelity in Fig.~\ref{fig:CPNN}(c) shows distinctly different behavior. At 0~km, there is very slight wavelength dependence. But as the channel length increases, so does the variation in relative process fidelity, as expected as PMD in the fiber grows.

For each of the four quantum channel lengths we tested, we performed QPT for several different combinations of quantum signal level and co-propagating classical networking laser power. Our base quantum signal level, $\alpha_0$, corresponds to a detected count rate of about 125 kcps at 0.5~km. $P_0$ corresponds to a classical networking laser launch power $P\approx-8.5$~dBm into the quantum channel. In Fig.~\ref{fig:CP15km}, we show the results for the 15.5-km quantum channel. The results for 0.5 and 5.5~km are in Appendix~\ref{App:QPTdata}. Figure~\ref{fig:CP15km} offers evidence of linear noise scaling, consistent with the dominant source being Raman scattering, the intensity of which is proportional to classical networking laser power~\cite{1253508,Peters_2009}: the unitarity, channel capacity, and relative process fidelity results are nearly identical for ($\alpha_0$,$P_0$) and ($2\alpha_0$,2$P_0$).

We see further characteristic signatures of Raman scattering in the overall shape of Fig.~\ref{fig:CP15km}(a). We would expect higher unitarity toward the classical networking laser wavelength (1542.5~nm) where there is less Raman scattering. Conversely, we expect lower unitarity on the higher wavelength side ($>$1555~nm) where the Raman scattering is slightly greater than the lower wavelength side ($<$1535~nm)~\cite{1253508,Peters_2009, Eraerds_2010}. When $P\neq0$, the highest unitarity in Fig.~\ref{fig:CP15km}(a) is found for ($2\alpha_0$,$P_0$). Furthermore, it is interesting to note that the lower unitarity examples also correspond to higher relative process fidelity as a function of wavelength [Fig.~\ref{fig:CP15km}(c)] because the quantum channel shows less wavelength dependence as growing background noise overtakes the effects of PMD. For ($\alpha_0$,$P_0$), we compare the quantum channels measured for different distances in Fig.~\ref{fig:CPqc912}. Here we find a strong correlation of decreasing unitarity and channel capacity as the channel length increases. This behavior is expected for Raman-dominated noise where the noise grows with the channel length until the classical light becomes sufficiently attenuated for the Raman noise to begin decreasing~\cite{PhysRevX.2.041010}. In Appendix~\ref{App:QPTdata}, we show representative Kraus operators for the 5.5-km data in Fig.~\ref{fig:CPqc912}.

\begin{figure}
   \centerline{\includegraphics[scale=0.6]{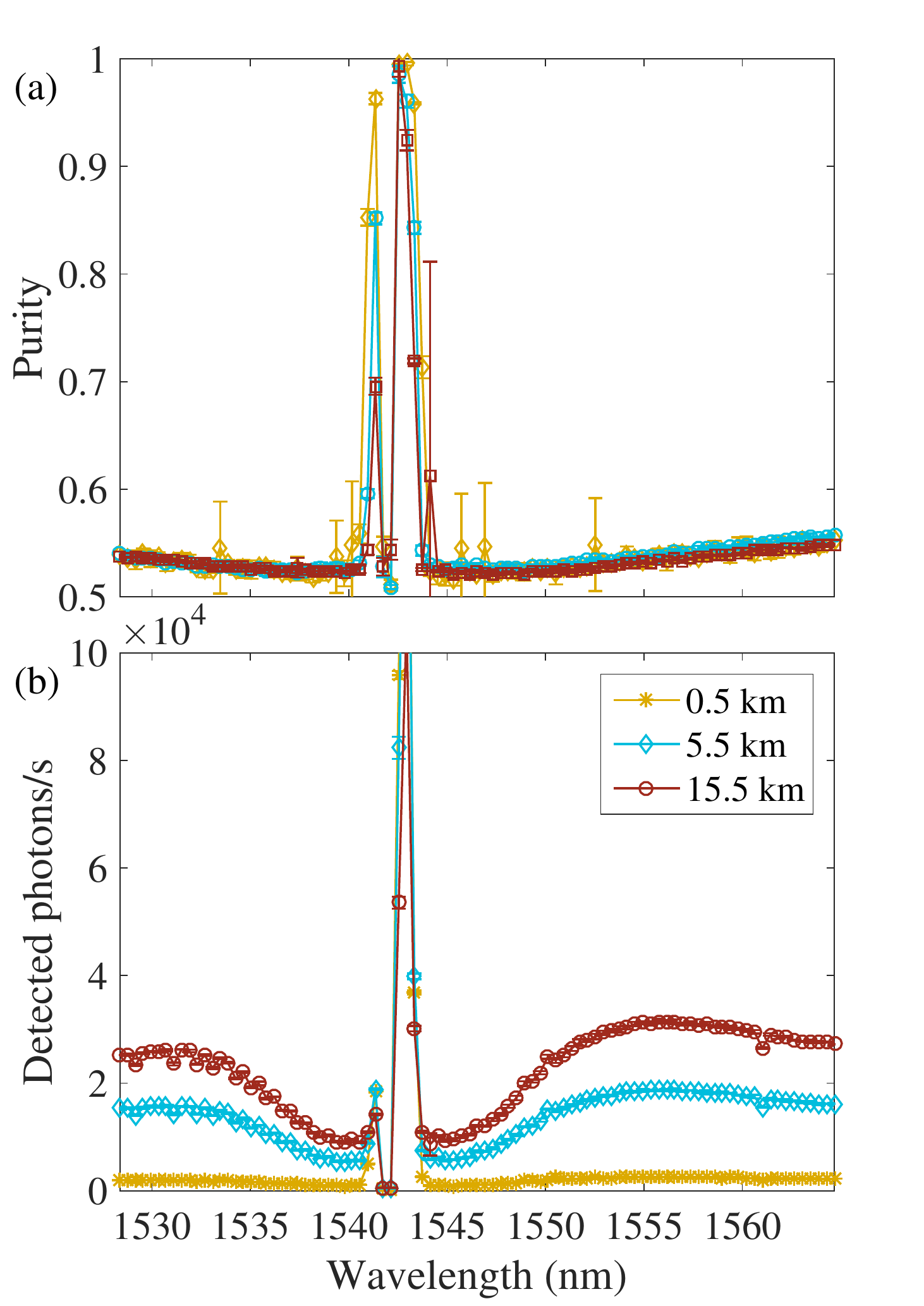}}
    \caption{Co-propagating noise characterization. (a) Average purity versus wavelength from five QST scans. (b) Average detected counts in H/V basis. The peaks near 1542.5 nm are due to leakage from the classical networking laser through the notch filter. The peak heights are about 177~kcps, 167~kcps, and 106~kcps for 0.5~km, 5.5~km, and 15.5~km, respectively. }
    \label{fig:copropnoise}
\end{figure}

For further insight, we turn off the quantum signal and perform QST of only the noise produced by the classical networking laser for 0.5~km, 5.5~km, and 15.5~km with $P=P_0$. We analyze the measurements using the Bayesian QST methods of Ref.~\cite{Lu2022b} and calculate state purity, $\mathcal{P}=\Tr\,\rho^2$, running five independent scans to gauge experimental uncertainty. Figure~\ref{fig:copropnoise}(a) shows that the state is partially polarized and close to the maximally mixed case of $\mathcal{P}=0.5$, which we discuss further in Sec.~\ref{sec:counter}. We also plot the average detected counts in the H/V basis in Fig.~\ref{fig:copropnoise}(b). For the longer fiber lengths, it is easy to see the expected shape of Raman scattering, which is low near the classical networking laser (1542.5~nm) and peaks, somewhat away from the classical networking laser, slightly higher on the longer wavelength side~\cite{Peters_2009,Eraerds_2010}. We can also see some artifacts from imperfect filtering near 1542~nm which allows some of the tails of the classical networking laser to be detected near 1542~nm, resulting in a rise in purity since the laser is polarized. Away from 1542~nm, the filter isolation is sufficiently high that there is no measurable classical networking laser leakage.

\subsection{Counter-propagating configuration}
\label{sec:counter}
\begin{figure}
   \centerline{\includegraphics[scale=0.6]{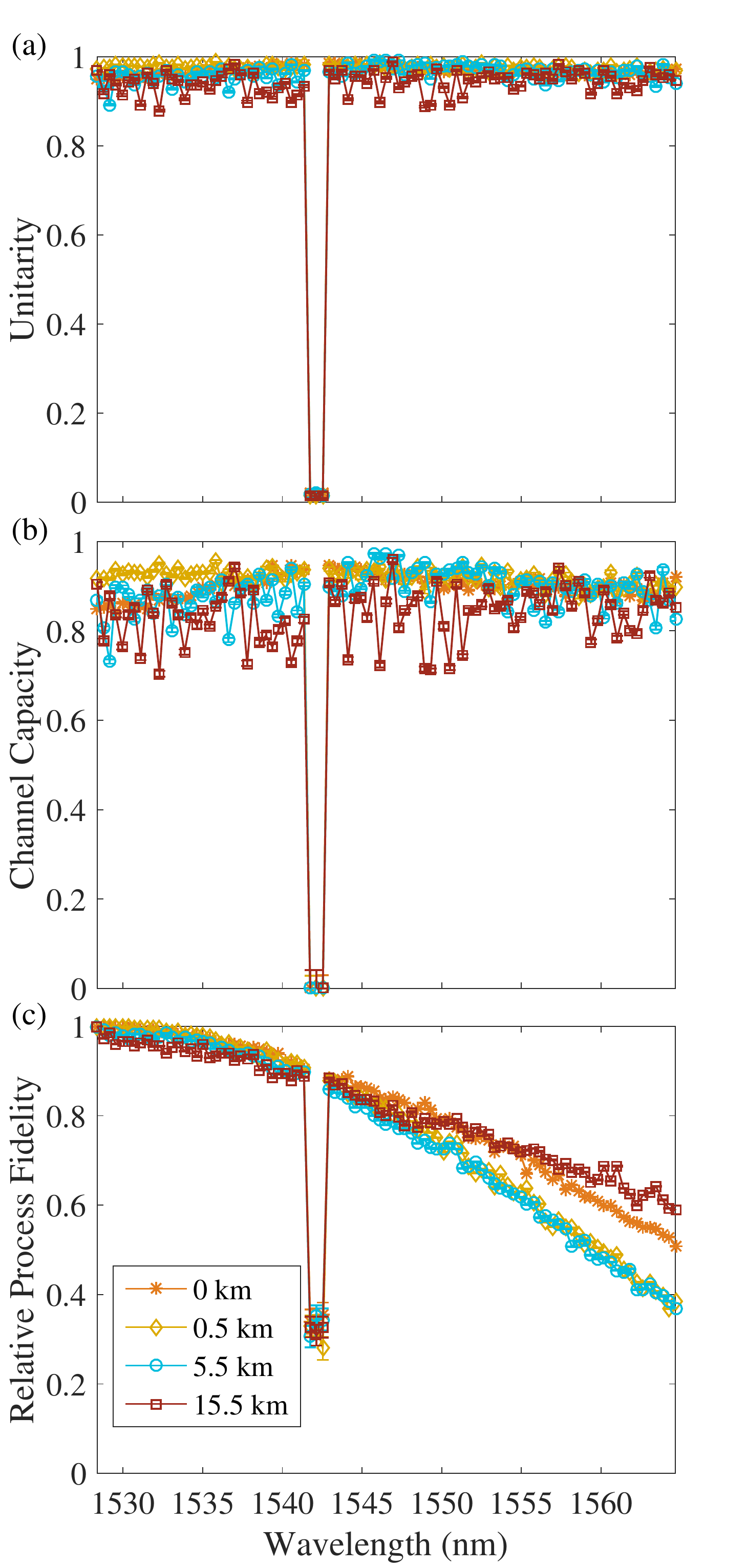}}
    \caption{Quantum process tomography with no added noise in counter-propagating configuration. (a) Unitarity,  (b) channel capacity (qubits/channel use), and (c) relative process fidelity versus wavelength for several different fiber lengths. We used the Kraus operators of the quantum channel at 1528.38~nm as the reference process in (c). The dip near 1542~nm results from the notch filter.}
    \label{fig:CNTRPNN}
\end{figure}

\begin{figure}
   \centerline{\includegraphics[scale=0.6]{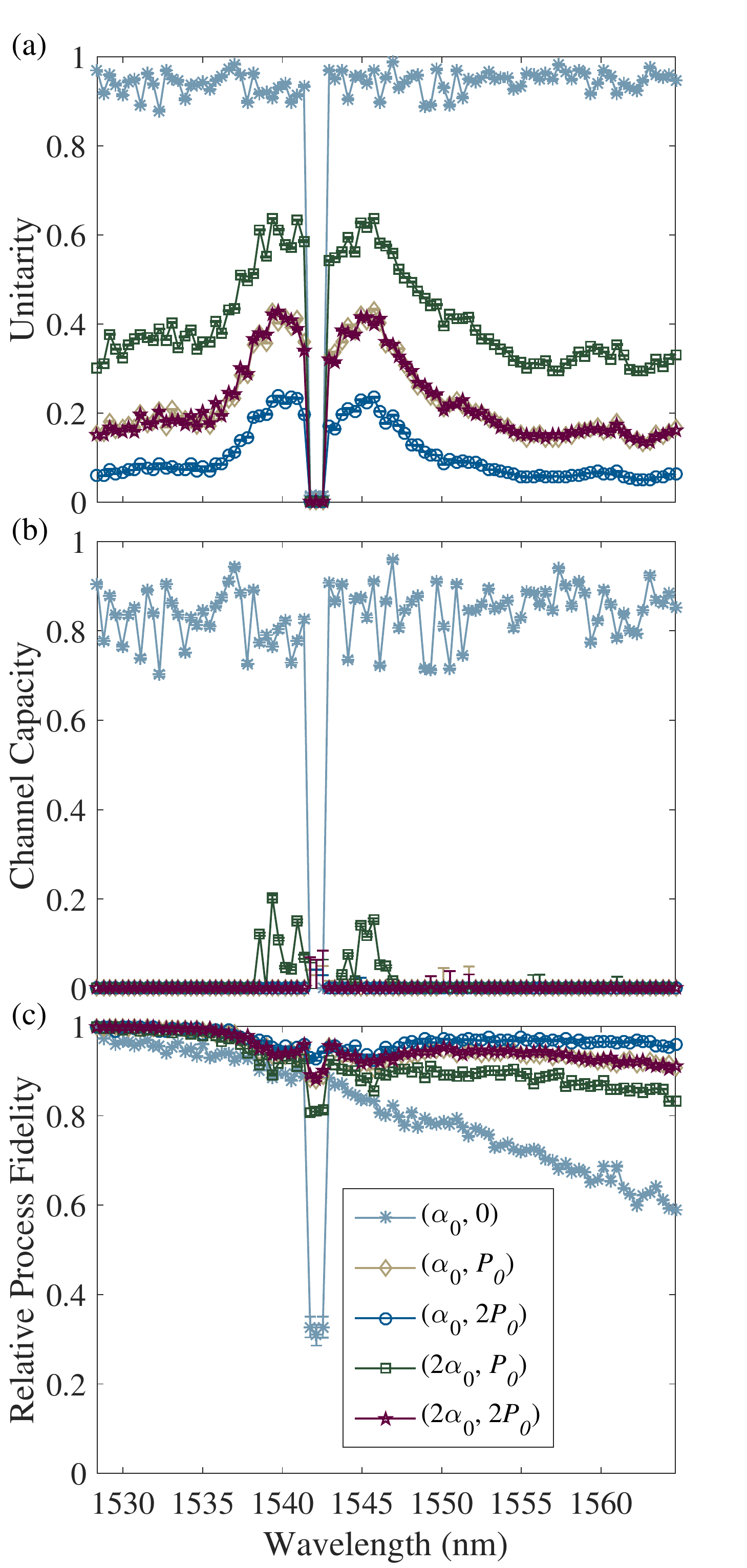}}
    \caption{Quantum process tomography at 15.5~km with added counter-propagating noise.  (a) Unitarity, (b) channel capacity (qubits/channel use), and (c) relative process fidelity versus wavelength for several different combinations of quantum signal and classical noise levels.  Our base quantum signal level, $\alpha_0\approx$125 kcps detected at 0.5~km. $P_0\approx-8.5$~dBm launch power of the classical networking laser into the quantum channel. We used the Kraus operators of the quantum channel at 1528.38~nm as the reference process in (c). The dip near 1542~nm is an artifact from the notch filter.}
    \label{fig:CNTRP15km}
\end{figure}

\begin{figure}
   \centerline{\includegraphics[scale=0.6]{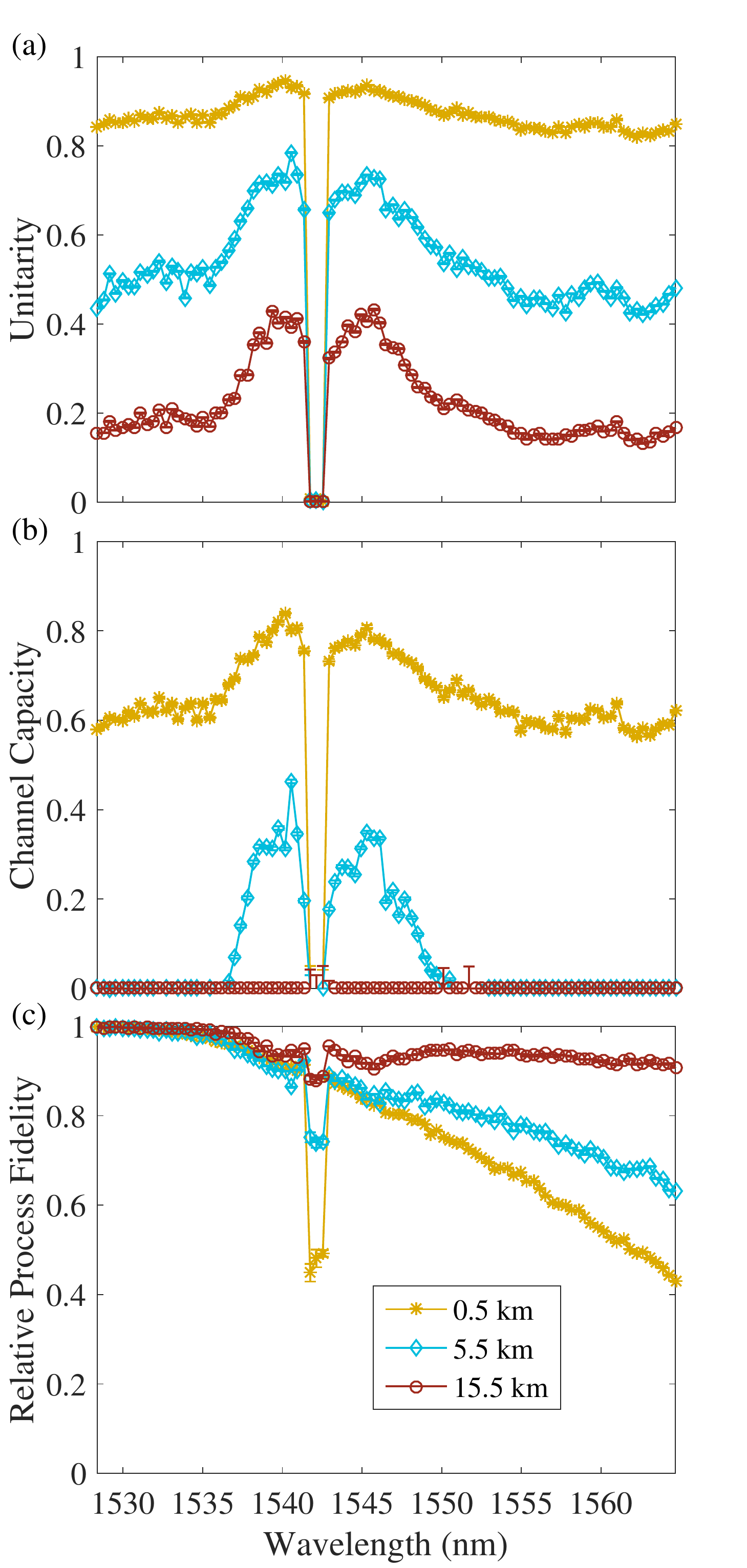}}
    \caption{Quantum process tomography versus wavelength for several distances with added counter-propagating noise ($\alpha_0$,$P_0$). (a) Unitarity, (b) channel capacity (qubits/channel use), and (c) relative process fidelity for a variety of fiber lengths. Our base quantum signal level, $\alpha_0\approx$125 kcps detected at 0.5~km. $P_0\approx-8.5$~dBm launch power of the classical networking laser into the quantum channel. We used the Kraus operators of the quantum channel at 1528.38~nm as the reference process in (c). The dip near 1542~nm is an artifact from the notch filter.}
    \label{fig:CNTRPqc912}
\end{figure}

Coexistence-induced impairments can occur when the quantum and classical signals travel in opposite directions as well; indeed, Raman backscattering flux is expected to exceed that of forward scattering for the same classical networking laser power and long fiber lengths~\cite{PhysRevX.2.041010}. Consequently, it is critical to examine both co- and counter-propagating configurations for a comprehensive evaluation of noise in fiber-optic links.
In Fig.~\ref{fig:CNTRPNN}, we show the results of our QPT scans for several different fiber lengths in the counter-propagating configuration, including a reference scan without added fiber. As before, in all panels of Fig.~\ref{fig:CNTRPNN}--\ref{fig:CNTRPqc912} the notch filter used in the optical system to remove backward scattering from the 1542.5-nm classical networking laser leads to a sharp drop in unitarity and capacity around 1542~nm. As with the co-propagating data in Fig.~\ref{fig:CPNN}(a), the counter-propagating data in Fig.~\ref{fig:CNTRPNN}(a) show stable unitarity of all fiber lengths measured when there is no added noise.  However, the relative process fidelity in Fig.~\ref{fig:CNTRPNN}(c) is significantly more wavelength-dependent than the corresponding co-propagating example. Because this dispersion occurs even in the 0.5-km case without added noise, we believe it results from the circulator used in the counter-propagating configuration in place of the 50/50 fused coupler (Fig.~\ref{fig:setup}).

For the 15.5-km quantum channel with noise from a counter-propagating classical laser, we perform QPT of different combinations of the quantum signal level and classical networking laser power (Fig.~\ref{fig:CNTRP15km}). The results for 0.5 and 5.5~km are in Appendix~\ref{App:QPTdata}. From Fig.~\ref{fig:CP15km}, we expect Raman scattering to be the dominant noise source,
and in agreement with this expectation see evidence of the linearity and the characteristic (inverted) Raman shape in Fig.~\ref{fig:CNTRP15km}(a). We do measure much lower unitarities and channel capacities with added noise compared to the co-propagating cases, while 
continuing to observe the correlation between lower unitarity and higher relative process fidelities [cf. Fig.~\ref{fig:CNTRP15km}(c)] due to the added noise overshadowing the effects of PMD. This feature is even more pronounced in the counter-propagating configuration since there is both more PMD from the circulator and higher Raman noise. For ($\alpha_0$,$P_0$), we compare the quantum channels measured in the counter-propagating configuration for different distances in Fig.~\ref{fig:CNTRPqc912}. Here we also see a strong correlation of decreasing unitarity and channel capacity as the channel length increases. 

\begin{figure}
   \centerline{\includegraphics[scale=0.6]{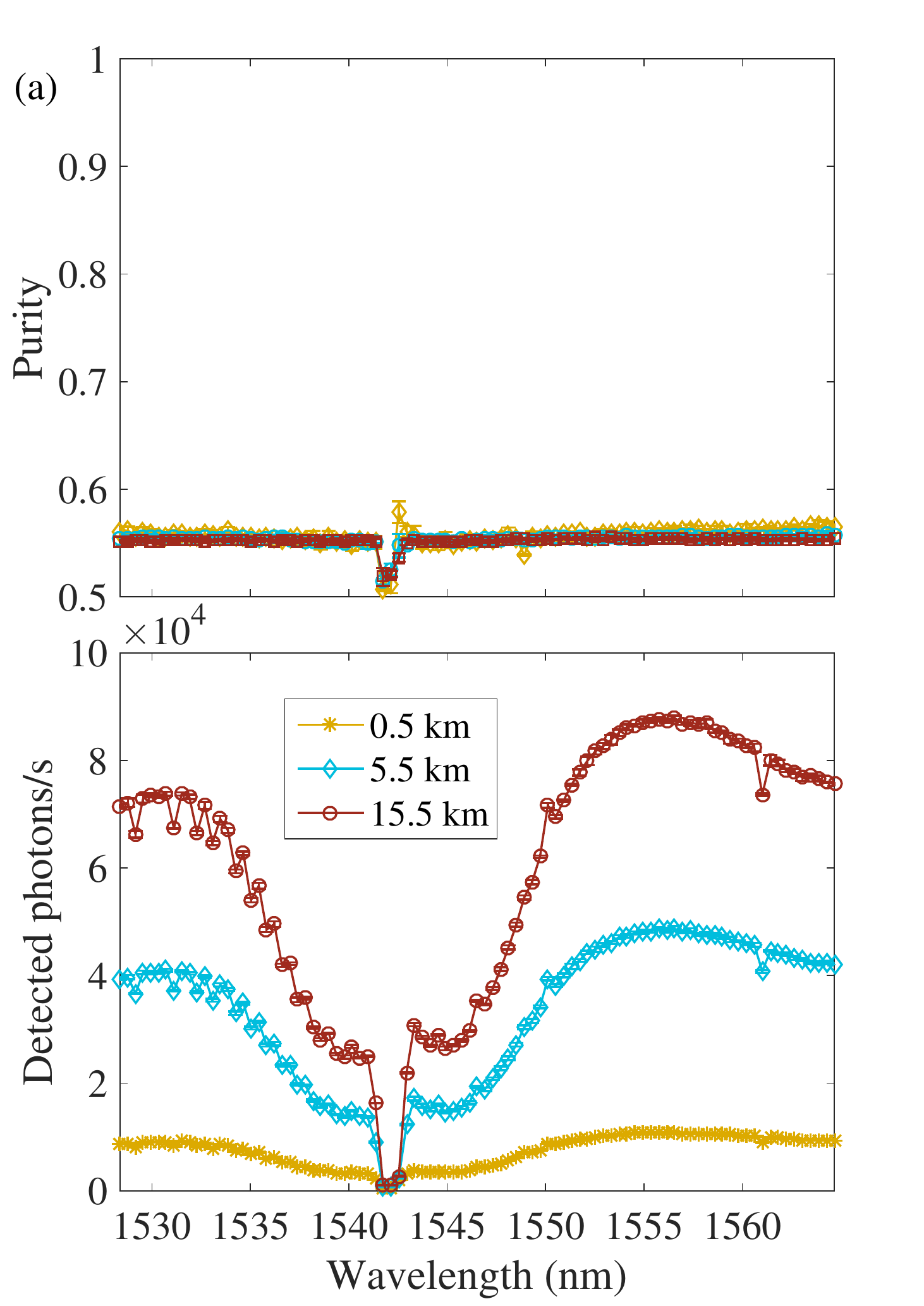}}
    \caption{Counter-propagating noise characterization. (a) Average purity versus wavelength from five QST scans. (b) Average detected counts in H/V basis.}
    \label{fig:counterpropnoise}
\end{figure}

Finally, we performed five full QST scans of only the noise produced by classical networking laser for 0.5~km, 5.5~km, and 15.5~km with $P=P_0$ in the counter-propagating configuration. Figure~\ref{fig:counterpropnoise}(a) plots the average purity from these scans.
We note that, unlike the co-propagating examples in Fig.~\ref{fig:copropnoise}(a), the purity does not rise near the notch filter since the polarized contributions from the laser propagate in the opposite direction. But like the co-propagating case, the noise state is highly mixed, although not maximally so ($\mathcal{P}=0.5$) due to slight imbalance in the $\ket{H}$ and $\ket{V}$ contributions. The polarization dependence of Raman gain is well known---and indeed a significant challenge for Raman amplifiers~\cite{stolen1979polarization,agrawal2001applications,Popov_2004}.
So while it is difficult to predict theoretically what the degree of polarization should be (due to unknowns in Raman gain and PMD), it is nonetheless unsurprising to observe partially polarized noise photons out of our fiber channel. 

In order to rule out the possibility of any systematic imbalance  from polarization-dependent loss (PDL) in our measurement apparatus, we mixed our classical networking laser (at half power) with an independent 1542.2-nm CW laser (Hewlett Packard 8168C) of equal power on a fiber-coupled polarizing beamsplitter (OZ Optics FOBS-22P-1111-9/125-SSSS-1550-PBS-60-3A3A3A3A-1-1) to produce an effectively unpolarized Raman pump. We again measured five QSTs of the noise at 5~km in the co-propagating and counter-propagating configurations and found the average quantum state purity to be 0.5008$\pm$0.0003 and 0.5010$\pm$0.0003, respectively, excluding the notch filter and nearby wavelengths (1541--1544~nm). These extremely low purities suggest negligible PDL in the measurement system, providing additional confirmation of true partial polarization in the results of Figs.~\ref{fig:copropnoise}(a) and \ref{fig:counterpropnoise}(a) due to polarization-dependent Raman gain. 

Finally, we plot the average detected counts in the H/V basis versus wavelength in Fig.~\ref{fig:counterpropnoise}(b). As the noise spectrum  expected for Raman scattering~\cite{Peters_2009, Eraerds_2010} appears in both Fig.~\ref{fig:copropnoise}(b) and Fig.~\ref{fig:counterpropnoise}(b), we conclude we do not have any significant amount of classical networking-laser leakage (except near 1542~nm in the co-propagating configuration). We also would not expect appreciable spontaneous four-wave mixing at these fiber lengths and have verified the same with other measurements by looking for coincidences between signal and idler wavelengths. Additionally, we find the back-scattered noise signal [Fig.~\ref{fig:counterpropnoise}(b)] to be 3--4$\times$ higher than the forward-scattered signal [Fig.~\ref{fig:copropnoise}(b)] instead of the same as each other~\cite{PhysRevX.2.041010} at these fiber lengths. This is most likely due to the high inhomogeneous loss throughout the inter-lab links; the 0.5-km link we used is deployed fiber on our campus, so these measurements represent realistic imperfections of a deployed system. Furthermore, it allows us to compare more variety of noise levels in the following section. 

\subsection{Comparison with canonical channel models}
Now we compare our measurements, specifically Fig.~\ref{fig:CPqc912} and Fig.~\ref{fig:CNTRPqc912},  with several known channel models. The depolarizing channel is a good candidate because, as confirmed in our tests above, Raman scattering produces (mostly) depolarized light. 
We also compare with the dephasing model as an alternative that is similar to the depolarizing model for small errors. Many quantum-channel formulations use a mixing probability $p_M$ to characterize the probability of passing the input state, $\rho$, unchanged or operating on it. For example, the depolarizing channel 
can be described in terms of Kraus operators as $\mathcal{E}_{\text{DePol}}(\rho)=(1-3p_M/4)\rho+(p_M/4)(\sigma_{\text{X}}\rho\sigma_{\text{X}}+\sigma_{\text{Y}}\rho\sigma_{\text{Y}}+\sigma_{\text{Z}}\rho\sigma_{\text{Z}})$, while the dephasing channel is represented as  $\mathcal{E}_{\text{DePh}}(\rho)=(1-p_M)\rho+p_M\sigma_{\text{Z}}\rho\sigma_{\text{Z}}$. Using these channel models, we calculate the process fidelity between the average measured Choi matrix over 1555--1560~nm for each distance and Choi matrices derived from the depolarizing and dephasing channel models. We numerically maximize our fidelity calculations over local unitary rotations to remove reference frame considerations.  Accordingly, the process fidelity allows us to compare our measured channel with a model that includes multiple distinct physical properties, e.g., noise injection and passive rotations. To calculate $p_M$ from the experimental data, we use the total counts (calculated using the H/V basis measurements) at a given wavelength for ($\alpha_0$,0) and (0,$P_0$) inputs, and average over 1555--1560~nm at each channel length.

\begin{figure}
    \centering
    \includegraphics[scale=0.6]{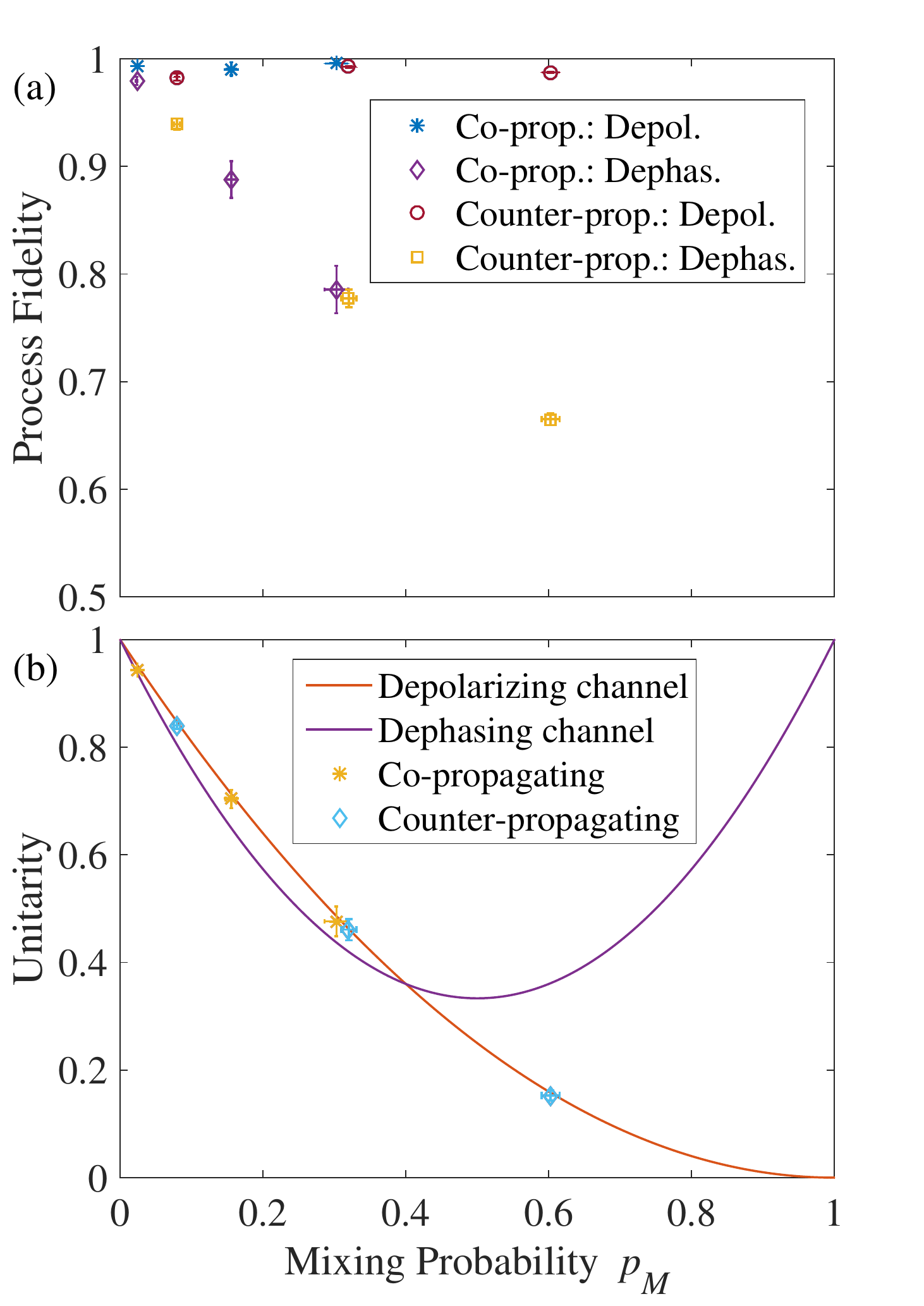}
    \caption{ Comparison with standard channel models. (a) Process fidelity between the rotated inferred Choi matrix for the ($\alpha_0$,$P_0$) datasets and that of depolarizing and dephasing models with mixing probability $p_M$. The Choi matrices derived from the data were averaged over 1555--1560~nm for each data point. We calculated $p_M$ using the total counts (calculated using the H/V basis measurements) at a given wavelength of ($\alpha_0$,0) and (0,$P_0$), averaged over 1555--1560~nm for each distance. The error bars represent one standard deviation. The co- and counter-propagating process fidelities with the depolarizing channel are all above 0.99 and 0.98, respectively. (b) Unitarity versus mixing probability for co- and counter-propagating ($\alpha_0$,$P_0$) datasets [Fig.~\ref{fig:CPqc912}(a) and Fig.~\ref{fig:CNTRPqc912}(a)], along with the dephasing and depolarizing channels.}
    \label{fig:PFunicomp}
\end{figure}

In Fig.~\ref{fig:PFunicomp}(a), we find very high fidelity ($>$0.98) between our measurements at all distances and the depolarizing channel, whereas the fidelity between our measurements and the dephasing channel decreases substantially with increasing $p_M$. This corroborates well with our direct measurements of the state of the noise [Fig.~\ref{fig:copropnoise}(a) and Fig.~\ref{fig:counterpropnoise}(a)], which although partially polarized are close to being maximally mixed. 
Finally, in Fig.~\ref{fig:PFunicomp}(b) we compare the unitarity of the depolarizing channel and dephasing channels to the unitarity of the ($\alpha_0$,$P_0$) datasets in Fig.~\ref{fig:CPqc912}(a) and Fig.~\ref{fig:CNTRPqc912}(a), averaged over 1555--1560~nm for each distance. The measured unitarities clearly follow the trend expected for a depolarizing channel, with appreciable deviation from a dephasing model at all values of $p_M$ but particularly strong for $p_M>0.4$. 

\begin{figure}
    \centering
    \includegraphics[scale=0.56]{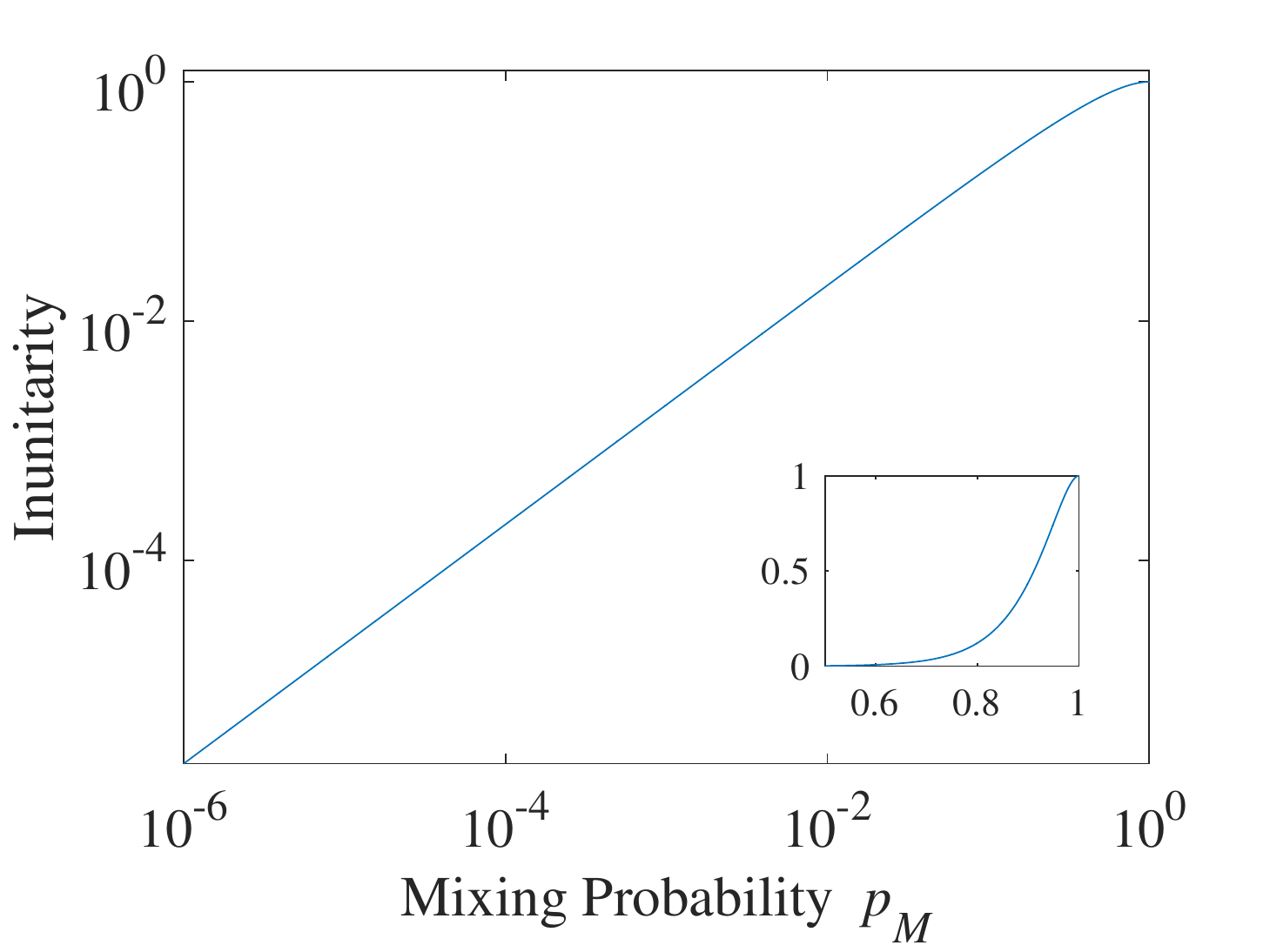}
    \caption{Inunitarity versus mixing probability for the depolarizing channel. Inset shows linearly scaled zoom of top right corner of main plot.}
    \label{fig:psim}
\end{figure}

For CW classical noise mixed with a pulsed quantum signal using detector gating or coincidence filtering, the mixing probability is proportional to the duty cycle of the detection window. Leveraging this fact, the mixing probability can be substantially decreased for a given fiber length. In Fig.~\ref{fig:psim}, we simulate the effect on the unitarity (plotted as ``inunitarity'' $1-U$) for decreasing mixing probability $p_M$. 
For the noise model we use the depolarizing channel due to the high process fidelity it has with the measured channel [Fig.~\ref{fig:PFunicomp}(a)]. In Fig.~\ref{fig:psim}, we see a linear relationship between the inunitarity and $p_M$ for small $p_M$, which provides favorable scaling for reducing the inunitarity by making $p_M$ smaller. In addition to decreasing $p_M$ using time filtering, spectral filtering can also be used, assuming there are not other factors limiting the unitarity. Using spectral filtering in addition to temporal filtering can allow for higher temporal duty cycles for the same $p_M$, which could allow for higher distribution rates at a fixed repetition rate.

\section{Conclusion}\label{section:conclusion}
Quantum and classical signals coexisting in the same fiber without quantum signal degradation is an important practical milestone for deploying large-scale quantum networks. Here we have systematically characterized the quantum channel produced by a classical networking laser for several fiber lengths and different propagation configurations. For this, we have developed and applied a Bayesian QPT reconstruction method combining an efficient MCMC algorithm and parametrization. 
Our results clearly show that the noise produced by classical networking lasers, predominantly Raman scattering, has a significant effect on the channel in high-noise (high-$p_M$) situations. Furthermore, we argue that this noise can be modelled effectively using a depolarizing channel; this can inform future development of quantum repeater theory and quantum error correction in quantum networking. For experimentalists and engineers, we quantitatively show a path forward to developing a low-noise coexistent quantum channel, using a depolarizing noise model, by calculating how much filtering, i.e., what $p_M$, is required for a desired level of channel unitarity, assuming the noise is the limiting factor.
Now that channel characterization has been done, we can move forward to coexistence demonstrations knowing the filtering requirements necessary to produce a low-noise quantum channel.
\section{Acknowledgments}
We acknowledge Brian P. Williams for his contribution to the development of the timetagger firmware. This work was performed at Oak Ridge National Laboratory, operated by UT-Battelle for the U.S. Department of Energy under contract no. DE-AC05-00OR22725. Funding was provided by the U.S. Department of Energy, Office of Science, Office of Advanced Scientific Computing Research, through the Transparent Optical Quantum Networks for Distributed Science Program and the Early Career Research Program (Field Work Proposals ERKJ355 and ERKJ353).

J.C.C., M.A., and N.R., constructed experimental setup and automated experiment control. J.C.C., M.A., and N.R. collected all measurements. J.C.C. and J.M.L. analyzed the data and ran simulations. J.C.C., J.M.L., and B.T.K developed Bayesian process tomography method and implemented the method in software. J.C.C, J.M.L, and N.A.P devised measurements and analysis for experiment. N.A.P supervised project and provided direction for measurements and analysis. All authors contributed to manuscript preparation.

\appendix

\section{Quantum Process Tomography Details} \label{App:QPT}
Beyond fundamental aspects of the algorithm itself, the speed of Bayesian inference with pCN depends heavily on the efficiency with which the likelihood can be calculated for any given parameter set. Following Ref.~\cite{Lu2022b}, we consider a Poissonian likelihood $L_D(\bx)$ of the form
\begin{equation}
\label{eq:LL}
L_\mathcal{D}(\bx) = \prod_{l=1}^L \prod_{j=1}^J e^{-\overline{N}_{lj}} \overline{N}_{lj}^{n_{lj}},  
\end{equation}
where $\overline{N}_{lj}=\Phi(z) \tau p_{l j}(\by)$ denotes the mean number of counts for input/output pair $\ket{\phi_l}/\ket{\psi_j}$, as defined after Eq.~\eqref{eq:plj}, but now explicitly showing dependencies on the underlying parameters. The vector $\bx\equiv (\by,z)$ includes both the $D^4$ complex numbers $y_j$ discussed in the main text for the quantum channel, and an additional parameter $z$ for flux $\Phi$. (See Ref.~\cite{Lu2022b} for details on this parametrization.) Finally, the set $\mathcal{D}=\{N_{11},...,N_{LJ}\}$ comprises the experimental counts observed for all settings---the data from which the quantum channel is to be inferred.

The main bottleneck centers on computation of the $LJ$ probabilities $p_{lj}$; with matrices $\psi_{m j}=\braket{m}{\psi_j}$ and $\phi_{n l}=\braket{n}{\phi_l}$, we can rewrite Eq.~\eqref{eq:plj} as
\begin{equation}
     p_{l j}=\sum_{k=1}^K|\psi_{m j}^*\phi_{n l}(A_k)_{m n}|^2.
\end{equation}
By defining the $L J \times D^2$ mapping matrix $M$ with elements
\begin{equation}
    M_{(l j)(m n)}=\psi_{m j}^* \phi_{n l}
\end{equation}
and viewing $(A_k)_{m n}$ as a $D^2$-dimensional vector, we have the matrix operation which computes the $[L J\times K]$-matrix
\begin{equation}
\label{eq:VMA}
    V=M(A_1 A_2 \cdots A_K),
\end{equation}
from which the probabilities follow as
\begin{equation}
    p_{l j}=\sum_{k=1}^K|V_{(l j)(k)}|^2.
\end{equation}

The mapping matrix $M$ depends only on the states selected for preparation and measurement, and therefore is fixed throughout the MCMC algorithm; the channel, however, as expressed by the $A_k$ matrices, varies with each sample $\by^{(s)}$, and so the efficiency of their calculation has a major impact on the total MCMC time. The approach we have adopted attempts to vectorize calculations when possible, and  begins by using the $y_j$ to populate, row by row, a single $K D\times D$ matrix comprising all $D\times D$ $G_k$ matrices:
\begin{equation}
    \mathbf{G}=\begin{pmatrix}
    G_1\\
    G_2\\
    \vdots\\
    G_K
    \end{pmatrix},
\end{equation}
from which follows the $D\times D$ matrix
\begin{equation}
    H=\sum_{k=1}^K {G_k^{\dagger}G_k}= \mathbf{G}^{\dagger}\mathbf{G} = \begin{pmatrix} G_1^{\dagger} & \cdots G_K^{\dagger}\end{pmatrix} \begin{pmatrix}  G_1\\
    \vdots\\
    G_K
    \end{pmatrix}.
\end{equation}
Then the Kraus operators are computed as
\begin{equation}
    \mathbf{A}=\begin{pmatrix}
    A_1\\
    \vdots\\
    A_K
    \end{pmatrix}=\begin{pmatrix}
    G_1\\
    \vdots\\
    G_K
    \end{pmatrix} H^{-1/2}
\end{equation}
and reshaped into a $D^2\times K$-matrix
\begin{equation}
    \mathbf{A'}= \begin{pmatrix} A_1^{'} & \cdots A_K^{'}\end{pmatrix},
\end{equation}
which is precisely the operator representation required in Eq.~\eqref{eq:VMA}.

For all MCMC computations, we keep $R=2^{10}$ samples for calculations from a longer chain of length $RT$, where the thinning factor $T$ is chosen to ensure convergence, found empirically by successively doubling $T$ from $T=2^2$ to $T=2^{18}$. As the chain was increased we expect the mean and standard deviation of any calculated quantity, e.g., unitarity, to converge to fixed values. We found the unitarity and its standard deviation to level off at $T=2^{10}$ for a representative measured dataset from Fig.~\ref{fig:CPNN}, and $T=2^{14}$ for ideal simulated data, i.e. Appendix~\ref{App:Chsim}. Thus for measured data we analyzed with $T=2^{11}$-length chains and for ideal simulations, we used $T^{15}$. After a similar test for quantum state tomography, we used $T=2^{10}$ for measured data.

To characterize the quantum channels reconstructed, we computed the unitarity~\cite{Wallman_2015, PhysRevResearch.4.023041} using the normalized identity and Pauli matrices as our orthonormal basis. The unitarity, which ranges from $[0,1]$, quantifies how closely the process can be described by a single unitary operation. The unitarity $U(\mathcal{E})=0$ if and only if $\mathcal{E}$ is a completely depolarizing channel, and $U(\mathcal{E})=1$ if and only if $\mathcal{E}$ is an isometry channel~\cite{PhysRevResearch.4.023041}. Specifically, we define the unitarity as 
\begin{equation}
    U(\mathcal{E})=\frac{1}{D^2-1}\Tr[T^{\dagger}T],
\end{equation}
where $T$ is a subsystem of $\mathcal{E}$ in an orthonormal basis of operators $\{X_{\mu}\}$. In this type of basis, $\mathcal{E}$ can be decomposed into
\begin{equation}
    \mathbf{\mathcal{E}}=\sum_{\mu=0}^{d^2-1}{\ket{\mathcal{E}(\mathbf{X_{\mu}})}\bra{\mathbf{X_{\mu}}}}=\begin{pmatrix} 1 & \mathbf{0}\\
    \mathbf{x} & T\\
    \end{pmatrix},
\end{equation}
where we use the Liouville representation $\mathcal{L}(\mathcal{E})$ by the relation $\mathcal{L}(\mathcal{E})\ket{\text{vec}(M)}=\ket{\text{vec}(\mathcal{E}(M))}$ and we use boldface to represent all vectorized quantities, i.e., $\ket{\mathbf{M}}\equiv\ket{\text{vec}(M)}$.

Furthermore, we computed the quantum channel capacity~\cite{PhysRevA.55.1613,PhysRevA.54.2629,Holevo_2020}, defined as
\begin{equation}
    \mathcal{C}(\mathcal{E})\equiv\underset{\rho}{\max} S(\mathcal{E}(\rho))-S_e(\rho,\mathcal{E}),
\end{equation}
where $S(\mathcal{E}(\rho))$ is the von Neumann entropy of $\mathcal{E}(\rho)$, $S_e(\rho,\mathcal{E})$ is the entropy exchange~\cite{PhysRevA.54.2614}, and the maximization (via sequential quadratic programming~\cite{nocedal2006numerical} in our case) is over all density operators $\rho$ which may be used as input to the channel. As defined in Ref.~\cite{PhysRevA.54.2614}, 
\begin{equation}
    S_e(\rho,\mathcal{E})=S(W),
\end{equation}
 where $W_{\mu\nu}=\Tr\,A_{\mu}\rho A_{\nu}$ and $A_\mu$ are the Kraus operators of the channel.

Finally, we characterize the wavelength dependence of the process matrices by the process fidelity. To do this, first we convert the channels, represented as Kraus operators, to Choi matrices following the method of Ref.~\cite{Wilde}
\begin{equation}
\mathfrak{C}\propto\begin{pmatrix}
\mathcal{E}(\ket{0}\bra{0}) & \dots & \mathcal{E}(\ket{0}\bra{d-1})\\
\vdots &  \ddots & \vdots\\
\mathcal{E}(\ket{d-1}\bra{0}) & \dots & \mathcal{E}(\ket{d-1}\bra{d-1})\\
\end{pmatrix},
\end{equation}
and normalize the matrix such that $\Tr\,\mathfrak{C} = 1$. Then we calculate the process fidelity of $\mathcal{E}_1$ and $\mathcal{E}_2$ using the generalized Ulhmann fidelity for any two positive semidefinite operators~\cite{Wilde}:
\begin{equation}
F_P(\mathfrak{C}_1,\mathfrak{C}_2)=\Bigg(\Tr \sqrt{\sqrt{\mathfrak{C}_1}\mathfrak{C}_2\sqrt{\mathfrak{C}_1}}\Bigg)^2.
\end{equation}

\section{Quantum Channel Simulations} \label{App:Chsim}
As simple examples of our Bayesian QPT reconstruction method described in Sec.~\ref{section:BQPT} and Appendix~\ref{App:QPT}, we simulate several canonical quantum channels by using the analytical channel form to generate noiseless simulated data of photon counts with an average rate of $10^5$ per integration time; we compute the unitarity of the resulting channel and compute the process fidelity with respect to the analytical channel.

One quantum channel that we simulate is the bit-flip channel~\cite{Wilde}, described by the transformation
\begin{equation}
    \mathcal{E}_{\text{BF}}(\rho)=(1-p_M)\rho+p_M\sigma_{\text{X}}\rho\sigma_{\text{X}}.
\end{equation}
This channel represents the case where  the computational basis states are flipped with probability $p_M$. Another simulated quantum channel is the dephasing channel~\cite{Wilde}, described by the transformation
\begin{equation}
    \mathcal{E}_{\text{DePh}}(\rho)=(1-p_M)\rho+p_M\sigma_{\text{Z}}\rho\sigma_{\text{Z}},
\end{equation}
which is aptly named since it describes processes where there is reduced phase coherence or there is phase averaging. This channel is also called the phase-flip channel because it describes a ``bit flip'' in the conjugate basis. The last simulated example is the depolarizing channel~\cite{Wilde}, described by the transformation
\begin{multline}
    \mathcal{E}_{\text{DePol}}(\rho)=\left(1-\frac{3p_M}{4}\right)\rho \\+ \frac{p_M}{4}\left(\sigma_{\text{X}}\rho\sigma_{\text{X}}+\sigma_{\text{Y}}\rho\sigma_{\text{Y}}+\sigma_{\text{Z}}\rho\sigma_{\text{Z}}\right).
\end{multline}
This is an example of a worst-case quantum process with incoherent error introduced by the channel. Another way to describe the channel is that with some probability we lose the input qubit and replace it with the maximally mixed state.

\begin{figure}
   \centerline{\includegraphics[scale=0.6]{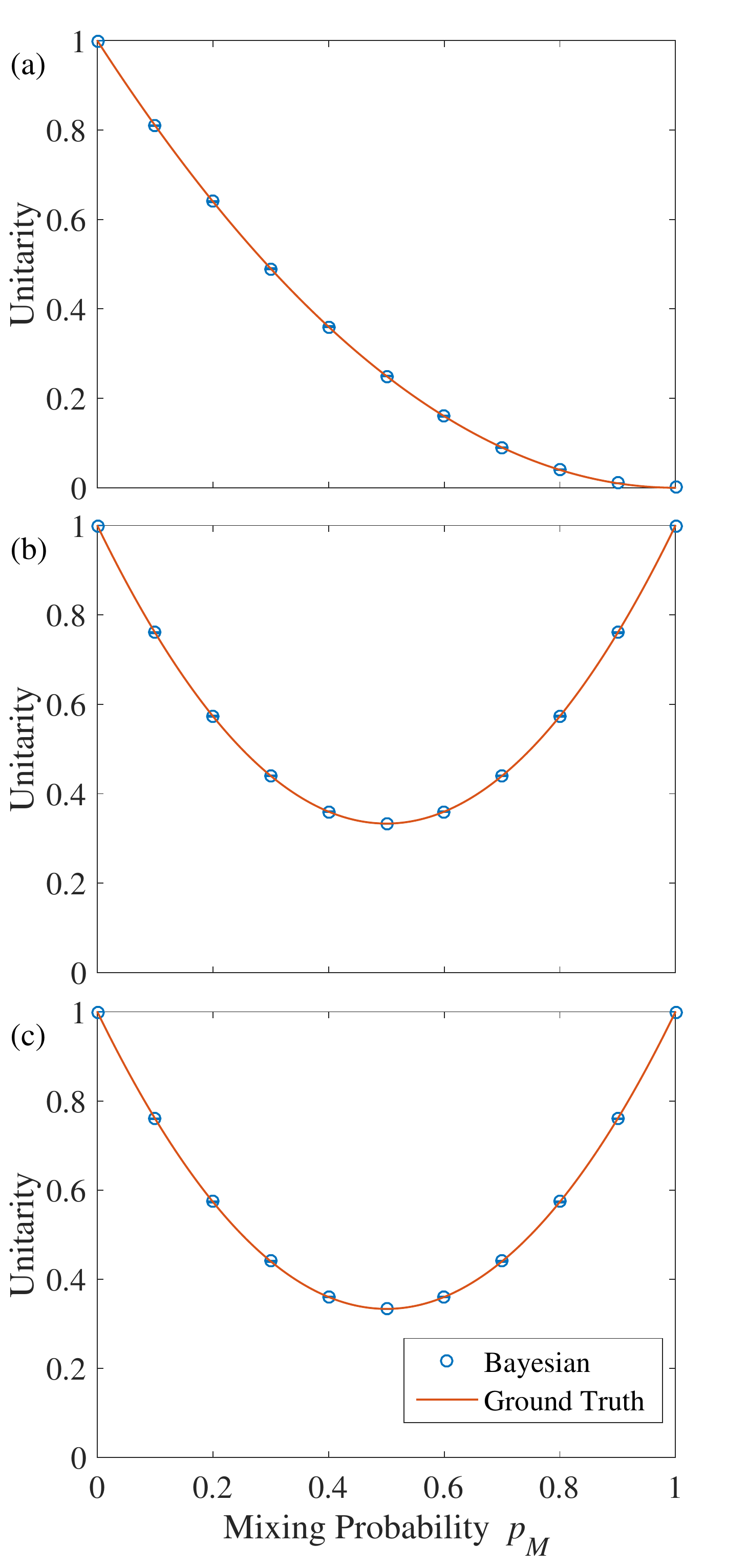}}
    \caption{Comparison between ground truth and Bayesian reconstruction from noiseless simulated data for the (a) depolarizing channel, (b) dephasing channel, and (c) bit-flip channel. The average difference of the reconstructed unitarities from the ground truth is $-7.8\times10^{-6}$, $-2.4\times10^{-5}$, and $-4.1\times10^{-5}$ for the depolarizing, dephasing, and bit-flip channels, respectively.}
    \label{fig:unisim}
\end{figure}

In Fig.~\ref{fig:unisim}, we plot the ground truth unitarity along with the unitarity of the Bayesian reconstruction 
and find excellent agreement between the two for all the channels. The average difference of the reconstructed unitarities from the ground truth is $-7.8\times10^{-6}$, $-2.4\times10^{-5}$, and $-4.1\times10^{-5}$ for the depolarizing, dephasing, and bit-flip channels, respectively. We also calculated the process fidelity between the ground truth Choi matrix of each simulated channel and the reconstructed Choi matrix from the Bayesian analysis to be 0.99998$\pm0.00001$, 0.99997$\pm0.00001$, and 0.99997$\pm0.00001$ for the depolarizing, dephasing, and bit-flip channels, respectively.

\section{Further Quantum Process Tomography Results} \label{App:QPTdata}
Here we show some representative Kraus operators from our measurements. For a distance of 5.5~km, using signal levels of ($\alpha_0$,$P_0$), at 1557 nm we reconstructed
\begin{equation}
    A_1=\begin{pmatrix}
    0.0381+0.4178i &-0.0898-0.2958i\\
    -0.0145-0.2147i & -0.0495-0.3312i
    \end{pmatrix}
\end{equation}
\begin{equation}
    A_2=\begin{pmatrix}
    0.0860-0.2309i &-0.3690+0.0941i\\
    -0.0796+0.2093i&-0.1682+0.3906i
    \end{pmatrix}
\end{equation}
\begin{equation}
    A_3=\begin{pmatrix}
    -0.4382-0.1713i &0.3781+0.2847i\\
    0.5199-0.2517i&0.3341-0.0378i
    \end{pmatrix}
\end{equation}
\begin{equation}
    A_4=\begin{pmatrix}
    -0.0222+0.0727i&-0.1800-0.2380i\\
    -0.1765-0.2737i&-0.0198-0.1995i
    \end{pmatrix}
\end{equation} in the co-propagating configuration, corresponding to $p_M=0.16\pm0.02$. In the counter-propagating configuration, corresponding to $p_M=0.32\pm0.02$, we reconstructed
\begin{equation}
    A_1=\begin{pmatrix}
    0.3938+0.0601i & -0.2140 -0.5590i\\
    -0.0747+0.1122i &-0.2601 - 0.1940i
    \end{pmatrix}
\end{equation}
\begin{equation}
    A_2=\begin{pmatrix}
    0.2898 - 0.1598i & -0.0887-0.4070i\\
    -0.1407+0.4613i &-0.4782 +0.1696i
    \end{pmatrix}
\end{equation}
\begin{equation}
    A_3=\begin{pmatrix}
-0.0569+0.2726i&-0.0948+0.0187i\\
0.1053+0.0883i & -0.2448-0.0936i
    \end{pmatrix}
\end{equation}
\begin{equation}
    A_4=\begin{pmatrix}
    0.1759+0.4439i &-0.0185-0.0263i\\
    -0.3698+0.1411i&0.0087-0.1625i
    \end{pmatrix}.
\end{equation}

\begin{figure}
   \centerline{\includegraphics[scale=0.6]{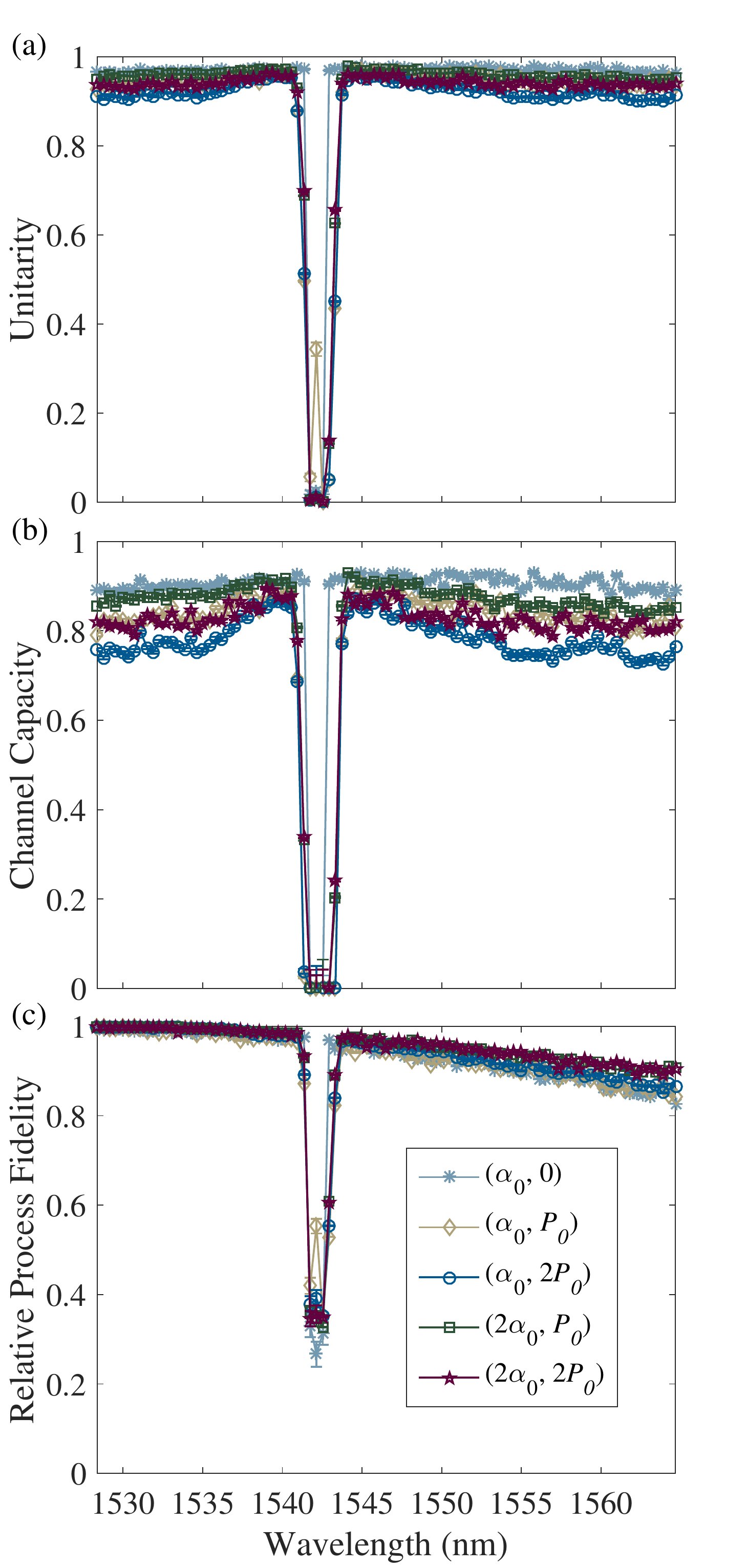}}
    \caption{Quantum process tomography at 0.5~km with added co-propagating noise.  (a) Unitarity, (b) channel capacity (qubits/channel use), and (c) relative process fidelity versus wavelength for several different combinations of quantum signal and classical noise levels.  Our base quantum signal level, $\alpha_0\approx$125 kcps detected at 0.5~km. $P_0\approx-8.5$~dBm launch power of the classical networking laser into the quantum channel. We used the Kraus operators of the quantum channel at 1528.38~nm as the reference process in (c). The dip near 1542~nm is an artifact from the notch filter.}
    \label{fig:CP0pt18km}
\end{figure}

\begin{figure}
   \centerline{\includegraphics[scale=0.6]{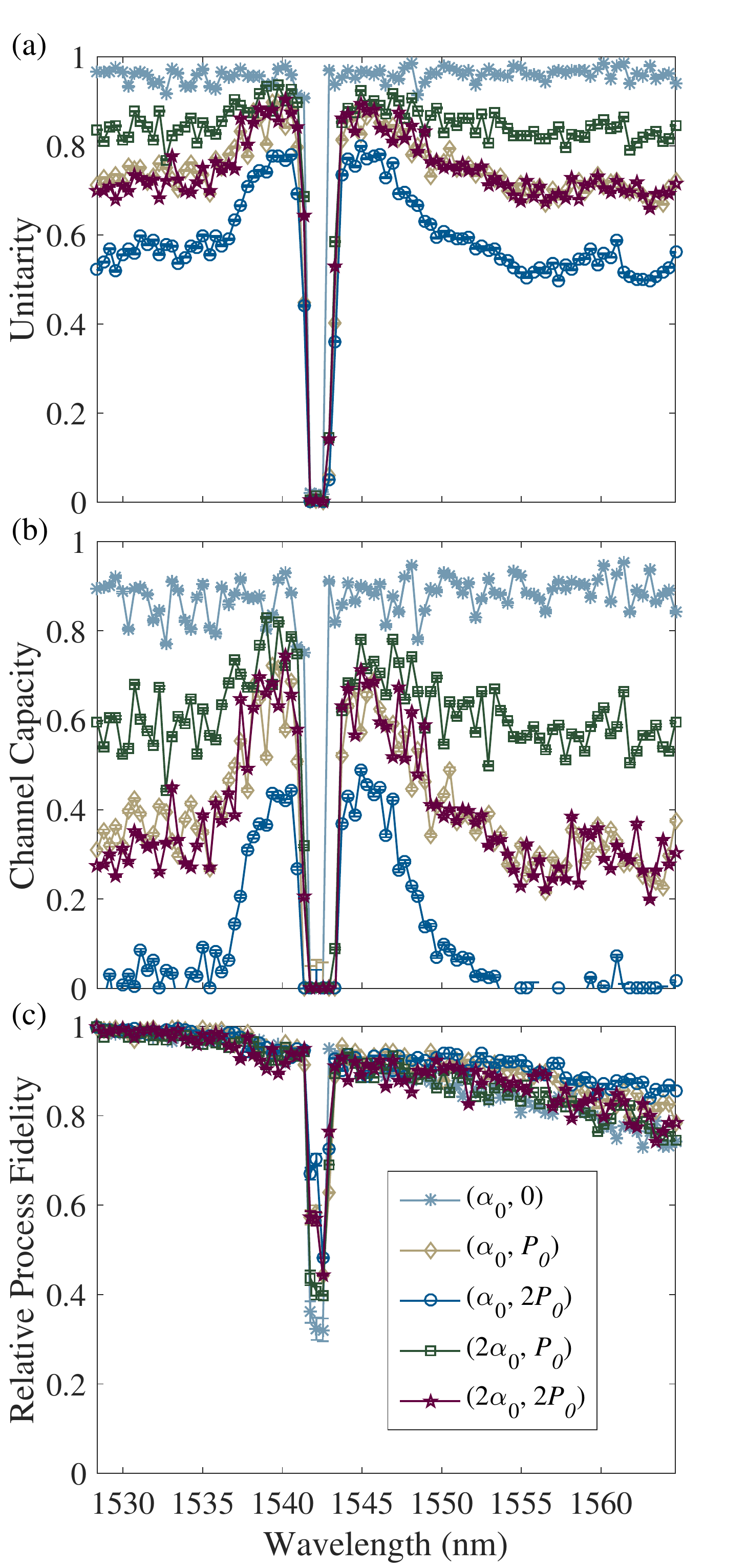}}
    \caption{Quantum process tomography at 5.5~km with added co-propagating noise.  (a) Unitarity, (b) channel capacity (qubits/channel use), and (c) relative process fidelity versus wavelength for several different combinations of quantum signal and classical noise levels.  Our base quantum signal level, $\alpha_0\approx$125 kcps detected at 0.5~km. $P_0\approx-8.5$~dBm launch power of the classical networking laser into the quantum channel. We used the Kraus operators of the quantum channel at 1528.38~nm as the reference process in (c). The dip near 1542~nm is an artifact from the notch filter.}
    \label{fig:CP5km}
\end{figure}

\begin{figure}
   \centerline{\includegraphics[scale=0.6]{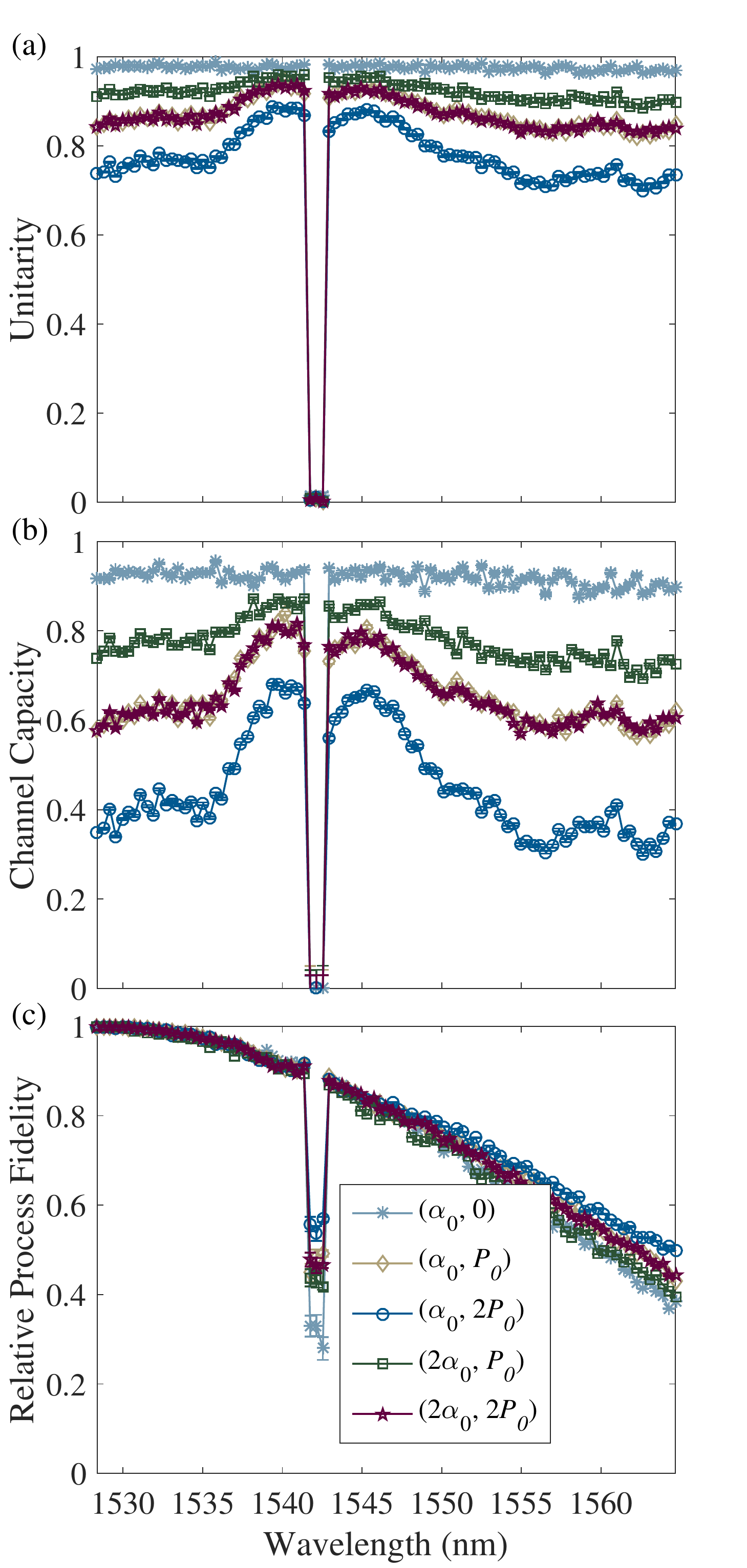}}
    \caption{Quantum process tomography at 0.5~km with added counter-propagating noise.  (a) Unitarity, (b) channel capacity (qubits/channel use), and (c) relative process fidelity versus wavelength for several different combinations of quantum signal and classical noise levels.  Our base quantum signal level, $\alpha_0\approx$125 kcps detected at 0.5~km. $P_0\approx-8.5$~dBm launch power of the classical networking laser into the quantum channel. We used the Kraus operators of the quantum channel at 1528.38~nm as the reference process in (c). The dip near 1542~nm is an artifact from the notch filter.}
    \label{fig:CNTRP0pt18km}
\end{figure}

\begin{figure}
   \centerline{\includegraphics[scale=0.6]{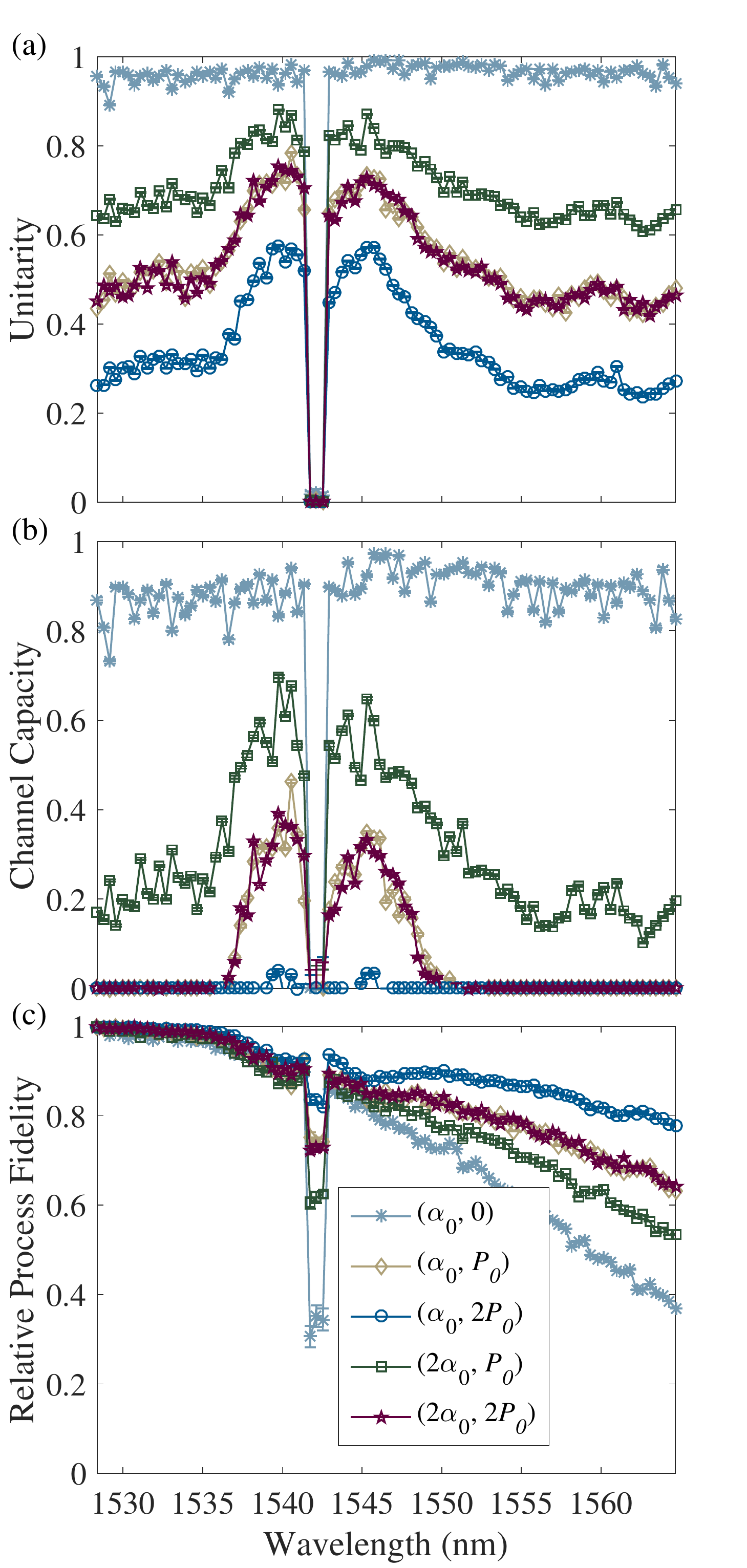}}
    \caption{Quantum process tomography at 5.5~km with added counter-propagating noise.  (a) Unitarity, (b) channel capacity (qubits/channel use), and (c) relative process fidelity versus wavelength for several different combinations of quantum signal and classical noise levels.  Our base quantum signal level, $\alpha_0\approx$125 kcps detected at 0.5~km. $P_0\approx-8.5$~dBm launch power of the classical networking laser into the quantum channel. We used the Kraus operators of the quantum channel at 1528.38~nm as the reference process in (c). The dip near 1542~nm is an artifact from the notch filter.}
    \label{fig:CNTRP5km}
\end{figure}

Finally, co-propagating results for 0.5 and 5.5~km are in Fig.~\ref{fig:CP0pt18km} and Fig.~\ref{fig:CP5km}, respectively.
Counter-propagating results for 0.5 and 5.5~km are in Fig.~\ref{fig:CNTRP0pt18km} and Fig.~\ref{fig:CNTRP5km}, respectively.

\clearpage
%

\end{document}